\documentclass[%
superscriptaddress,
showpacs,preprintnumbers,
 amsmath,amssymb,
 aps,
prl,
twocolumn,
10pt,
]{revtex4-1}

\usepackage[pdftex]{hyperref,color}
\usepackage{colortbl}
\usepackage{units}
\usepackage{upgreek}
\usepackage{textcomp}

\usepackage{graphicx}
\usepackage{bm}
\usepackage[ansinew]{inputenc}

\begin{document}

\title{Europium Underneath Graphene on Ir(111):\texorpdfstring{\\}{} Intercalation Mechanism, Magnetism, and Band Structure}


\author{Stefan Schumacher}
\affiliation{II. Physikalisches Institut, Universit\"{a}t zu K\"{o}ln, Z\"{u}lpicher Stra\ss e 77, 50937 K\"{o}ln, Germany}
\author{Felix Huttmann}
\affiliation{II. Physikalisches Institut, Universit\"{a}t zu K\"{o}ln, Z\"{u}lpicher Stra\ss e 77, 50937 K\"{o}ln, Germany}
\author{Marin Petrovi\'{c}}
\affiliation{Institut za fiziku, Bijeni\v{c}ka 46, 10000 Zagreb, Croatia}
\author{Christian Witt}
\affiliation{Department of Physics and CENIDE, Universit\"{a}t Duisburg-Essen, Lotharstra\ss e 1, 47048 Duisburg, Germany}
\author{Daniel F. F{\"o}rster}
\affiliation{II. Physikalisches Institut, Universit\"{a}t zu K\"{o}ln, Z\"{u}lpicher Stra\ss e 77, 50937 K\"{o}ln, Germany}
\author{Chi Vo-Van}
\affiliation{Institut N\'{e}el, CNRS et Universit\'{e} Joseph Fourier, BP 166, 38042 Grenoble Cedex 9, France}
\author{Johann Coraux}
\affiliation{Institut N\'{e}el, CNRS et Universit\'{e} Joseph Fourier, BP 166, 38042 Grenoble Cedex 9, France}
\author{Antonio J. Mart\'{i}nez-Galera}
\affiliation{II. Physikalisches Institut, Universit\"{a}t zu K\"{o}ln, Z\"{u}lpicher Stra\ss e 77, 50937 K\"{o}ln, Germany}
\author{Violetta \surname{Sessi}}
\affiliation{European Synchrotron Radiation Facility, BP 220, 38043 Grenoble Cedex, France}
\author{Ignacio Vergara}
\affiliation{II. Physikalisches Institut, Universit\"{a}t zu K\"{o}ln, Z\"{u}lpicher Stra\ss e 77, 50937 K\"{o}ln, Germany}
\author{Reinhard R\"{u}ckamp}
\affiliation{II. Physikalisches Institut, Universit\"{a}t zu K\"{o}ln, Z\"{u}lpicher Stra\ss e 77, 50937 K\"{o}ln, Germany}
\author{Markus Gr\"{u}ninger}
\affiliation{II. Physikalisches Institut, Universit\"{a}t zu K\"{o}ln, Z\"{u}lpicher Stra\ss e 77, 50937 K\"{o}ln, Germany}
\author{Nicolas Schleheck}
\affiliation{II. Physikalisches Institut, Universit\"{a}t zu K\"{o}ln, Z\"{u}lpicher Stra\ss e 77, 50937 K\"{o}ln, Germany}
\author{Frank Meyer zu Heringdorf}
\affiliation{Department of Physics and CENIDE, Universit\"{a}t Duisburg-Essen, Lotharstra\ss e 1, 47048 Duisburg, Germany}
\author{Philippe Ohresser}
\affiliation{Synchrotron SOLEIL, L'Orme des Merisiers, BP48, Saint-Aubin, 91192 Gif-sur-Yvette, France}
\author{Marko Kralj}
\affiliation{Institut za fiziku, Bijeni\v{c}ka 46, 10000 Zagreb, Croatia}
\author{Tim O. Wehling}
\affiliation{Institut für Theoretische Physik, Universit\"{a}t Bremen, Otto-Hahn-Allee 1, 28359 Bremen, Germany}
\affiliation{BCCMS, Universit{\"a}t Bremen, Am Fallturm 1a, 28359 Bremen, Germany}
\author{Thomas Michely}
\affiliation{II. Physikalisches Institut, Universit\"{a}t zu K\"{o}ln, Z\"{u}lpicher Stra\ss e 77, 50937 K\"{o}ln, Germany}

\date{\today}

\hyphenation{pho-to-emis-sion}
\hyphenation{me-ta-mag-net-ic}
\begin{abstract}
The intercalation of Eu underneath Gr on Ir(111) is comprehensively investigated by microscopic, magnetic, and spectroscopic measurements, as well as by density functional theory.
Depending on the coverage, the intercalated Eu atoms form either a $(2 \times 2)$ or a $(\sqrt{3} \times \sqrt{3})$R$30^{\circ}$ superstructure with respect to Gr.
We investigate the mechanisms of Eu penetration through a nominally closed Gr sheet and measure the electronic structures and magnetic properties of the two intercalation systems.
Their {electronic structures are} rather similar. 
Compared to Gr on Ir(111), the Gr bands in both systems are essentially rigidly shifted to larger binding energies resulting in n-doping.
The hybridization of the Ir surface state $S_1$ with Gr states is lifted, and the moiré superperiodic potential is strongly reduced.
In contrast, the magnetic behavior of the two intercalation systems differs substantially as found by X-ray magnetic circular dichroism.
The $(2 \times 2)$ Eu structure displays plain paramagnetic behavior, whereas for the $(\sqrt{3} \times \sqrt{3})$R$30^{\circ}$ structure the large zero-field susceptibility indicates ferromagnetic coupling, despite the absence of hysteresis at \unit[10]{K}.
For the latter structure, a considerable easy-plane magnetic anisotropy is observed and interpreted as shape anisotropy. 
\end{abstract}

\pacs{68.65.Pq, 71.20.Tx, 78.20.Ls, 75.70.-i}



\maketitle

\section{Introduction}
Graphene (Gr) is a promising material for spintronics and related applications because of its small spin-orbit and hyperfine interactions \cite{Min2006, Yazyev2008}. To make full use of this potential it is necessary to bring Gr in contact with magnetic materials. However, the ferromagnetic transition metals Fe, Co, and Ni possess localized valence orbitals of d symmetry that interact with the C p$_z$ orbitals and thereby destroy the electronic $\pi$ system of Gr \cite{Varykhalov2009, Weser2011}.

As an alternative, it was proposed to bring Gr in contact with the ferromagnetic insulator EuO  in order to induce an exchange-split Dirac cone \cite{Haugen2008}. In this context, it has recently been shown that high-quality EuO films can be grown on Gr on Ir(111) \cite{Klinkhammer2013, Klinkhammer2014}, and it would certainly be rewarding to analyze spin injection through EuO contacts into Gr.

As a further approach to induce a spin-split Dirac cone one might be tempted to bring Gr into contact with a magnetic metal that does not destroy the Gr band structure. In this respect rare earth elements are of interest as (i) most of them do not possess d orbitals close to the Fermi level that tend to form covalent bonds, and (ii) the magnetic moment carried by the highly localized 4f shell is very robust to the chemical environment. Questions that arise are: Is there a hybridization between the 4f states and Gr? Is there potential for ferromagnetic order of the 4f moments and is it possible to induce a spin-split Dirac cone? 

As a step of research in this direction we report here on the intercalation of Eu underneath Gr on Ir(111). We investigate the atomic structure by scanning tunneling microscopy (STM) and low-energy electron diffraction (LEED). Furthermore, we follow the intercalation process in real time by low-energy and photoemission electron microscopy (LEEM/PEEM). The magnetic properties are measured by X-ray magnetic circular dichroism (XMCD). Finally, we investigate the effect of intercalation on the Gr band structure by angle-resolved photoemission spectroscopy (ARPES) and relate this to Raman spectroscopic data. Our experimental results are complemented by density functional theory (DFT) calculations.

There are a number of points making Eu an interesting candidate for our study:
(i) Eu has a large magnetic moment of  $\unit[7]{\mu_{\mathrm{B}}}$ and displays  magnetic order below 90~K \cite{Olsen1962, Nereson1964, Arnold1964}.
(ii) The Eu binding to Gr is predominantly of ionic character and thus expected to leave the Gr band structure intact \cite{Forster2012b}.
(iii) Based on our experience with Eu intercalation \cite{Schumacher2013a, Schumacher2013b}, we are able to establish two different, well-defined intercalation phases of Eu underneath Gr on Ir(111), namely a $(2 \times 2)$ and a $(\sqrt{3} \times \sqrt{3})$R$30^{\circ}$ structure with respect to Gr. This opens the possibility to study the influence of the interatomic distance on the magnetic properties.
(iv) The $(\sqrt{3} \times \sqrt{3})$R$30^{\circ}$ structure also occurs in the layers of the well-investigated first-stage Eu graphite intercalation compound (GIC) EuC$_6$, which shows a complex metamagnetic behavior, i.e., the magnetic ordering changes with the magnitude of the magnetic field \cite{Suematsu1983, Chen1986}. Thus, Eu is well suited to investigate the changes in magnetism when going from the bulk Eu GIC to a single intercalation layer and to analyze the influence of the supporting substrate.
(v) The reactive Eu layer is efficiently protected underneath Gr against oxidation.
Thus, one may consider the investigation of the magnetic Eu layer under Gr also as unique chance to analyze the magnetic properties of a clean, unreacted, monatomic Eu layer, an endeavor that is practically impossible without the Gr cover.

\section{Experimental} 

As substrate we used an Ir(111) single crystal which was prepared by cycles of sputtering at \unit[920]{K} and annealing to \unit[1520]{K}, yielding clean terraces with sizes in the order of \unit[100]{nm}. A well oriented and perfectly closed Gr monolayer was prepared by room temperature ethylene adsorption till saturation, thermal decomposition at \unit[1500]{K}, and subsequent exposure to $\unit[5 \times 10^{-7}]{mbar}$ ethylene at \unit[1170]{K} for \unit[600]{s} (TPG+CVD method) \cite{vanGastel2009}.

High-purity Eu \cite{Ames} was evaporated from a water-cooled Knudsen cell at \unit[720]{K} sample temperature. Prolonged degassing of the Eu, which usually has a high H$_2$ content, ensured a background pressure below $\unit[1 \times 10^{-10}]{mbar}$ during growth. Using a quartz crystal microbalance, the Eu deposition rate was calibrated. We used rates  on the order of \unit[1]{\AA/min} equivalent to a Eu flux of $f_{\mathrm{Eu}}=\unit[3.5 \times 10^{16}]{atoms~m^{-2}s^{-1}}$.

Characterization of the intercalated structures was performed in an ultra-high vacuum (UHV) variable-temperature STM system in Cologne with base pressure of $\unit[3 \times 10^{-11}]{mbar}$. Images were taken  at room temperature and digitally post-processed with the WSxM software \cite{Horcas2007}. LEED measurements were performed using a three-grid rearview analyzer. 

Magnetic measurements by means of XMCD were conducted at the ID08 beamline of the European Synchrotron Radiation Facility (ESRF) in the total electron yield (TEY) mode using (99$\pm$1)\% circularly polarized light and external magnetic fields up to $\unit[5]{T}$.
While the magnetic field direction is always fixed either parallel or antiparallel to the direction of the incident light beam,
the sample can be rotated to adjust the angle $\theta$ of the surface normal to the incoming X-ray beam and external magnetic field. This allows us to obtain information about magnetic anisotropy.
In order to avoid non-magnetic artifacts resulting from switching of either the X-ray helicity or the external field direction, spectra were always measured for all four combinations of magnetic field direction and helicity and appropriately averaged. For brevity, we refer to X-ray helicity parallel (antiparallel) to the magnetic field simply as positive (negative) helicity.
The samples were prepared \emph{in situ} using the same preparation described above. The sample quality was verified by STM and LEED.

LEEM and PEEM measurements were performed using a commercial SPE-LEEM system \cite{Schmidt1998} in Duisburg. For PEEM experiments the sample was illuminated with light from a Hg discharge lamp ($h\nu=\unit[4.9]{eV}$) under $74^{\circ}$ grazing incidence.

Photoemission experiments were conducted in an ARPES-dedicated setup in Zagreb. A Scienta SES100 hemispherical electron analyzer (\unit[25]{meV} energy resolution, $0.2^{\circ}$ angular resolution) was used for data acquisition in two directions: along $\Gamma$KM and perpendicular to $\Gamma$KM. For excitation, a helium discharge lamp provided photons of \unit[21.2]{eV} energy with mixed polarization. The spectra were recorded at \unit[200]{K} ($\Gamma$KM) and \unit[250]{K} (perpendicular to $\Gamma$KM). 

Spin-polarized DFT calculations were performed using the generalized gradient approximation (GGA) \cite{Perdew1996} and  the projector augmented wave (PAW) basis sets \cite{Blochl1994, Kresse1993} as implemented in the Vienna Ab Initio Simulation Package (VASP) \cite{Kresse1994}. The strong local Coulomb interaction of the Eu 4f electrons was taken into account within the GGA+U approach using the Coulomb parameters $U = \unit[7]{eV}$ and $J = \unit[1]{eV}$ which have previously been shown to be well suited to describe rare earth systems \cite{Anisimov1997, Larson2007}.

\section{Structure of intercalated Europium}

As previously shown in Ref.~\onlinecite{Schumacher2013a}, Eu deposited at \unit[720]{K} intercalates under Gr on Ir(111). For partial coverages a complex pattern consisting of stripes, compact islands, and channels is formed. The pattern formation can be explained by the interplay of the chemically modulated binding of Gr to the substrate and the release of preexisting strain in Gr. Upon depositing more Eu, the layer can be completely filled. Only occasionally narrow channels and point-like spots at pentagon-heptagon defects are found where no Eu is intercalated.
\begin{figure} [htbp]
\begin{center}
\includegraphics[width=1.0\linewidth]{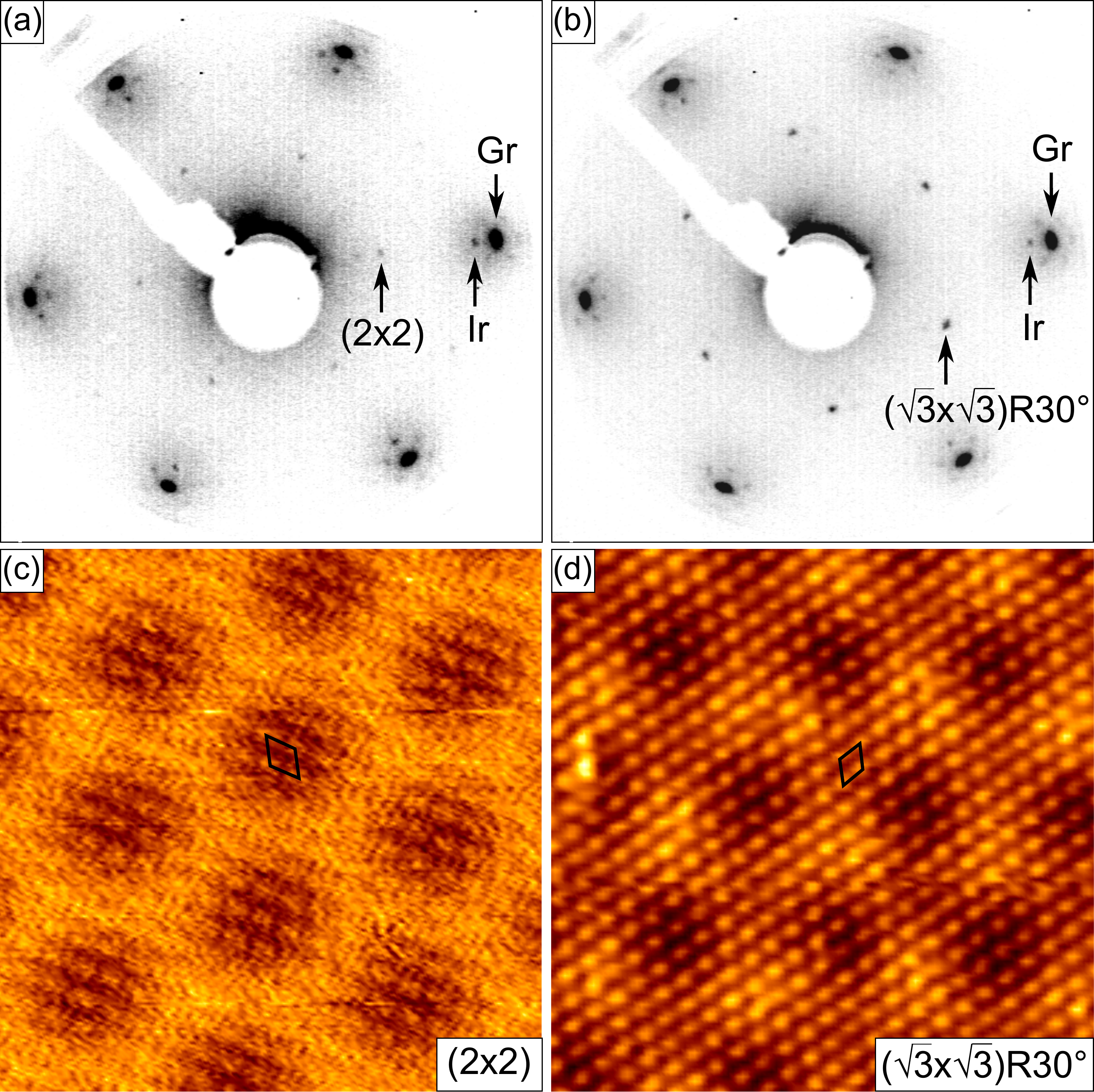}	
\caption{\label{figure1}
(color online) (a), (b) Inverted contrast LEED patterns at \unit[82]{eV} primary electron energy of a fully intercalated $(2 \times 2)$ and a $(\sqrt{3} \times \sqrt{3})\mathrm{R}30^{\circ}$ Eu layer, respectively.
First-order spots of the Ir substrate, the Gr layer, and the Eu superstructure are indicated. (c) STM topograph with superstructure resolution of the $(2 \times 2)$ Eu layer  ($I_{\mathrm{t}}=\unit[0.1]{nA}$, $U_{\mathrm{s}}=\unit[-62]{mV}$, $\unit[8.5]{nm} \times \unit[8.5]{nm}$).
The unit cell of the intercalated Eu is indicated by a diamond.
(d) STM topograph with superstructure resolution of the $(\sqrt{3} \times \sqrt{3})\mathrm{R}30^{\circ}$ Eu structure ($I_{\mathrm{t}}=\unit[1]{nA}$, $U_{\mathrm{s}}=\unit[-540]{mV}$, $\unit[8.5]{nm} \times \unit[8.5]{nm}$). Again, the unit cell of the intercalated Eu is indicated by a diamond.}
\end{center}
\end{figure} 

The intercalated Eu atoms adsorb in a $(2 \times 2)$ superstructure with respect to Gr as obvious in the LEED pattern in Fig.~\ref{figure1}(a). Depositing more Eu than the $(2 \times 2)$ saturation coverage (one Eu atom per four Gr unit cells) compresses the layer into a $(\sqrt{3} \times \sqrt{3})\mathrm{R}30^{\circ}$ structure with respect to Gr. The corresponding LEED pattern for a saturated layer (one Eu atom per three Gr unit cells) is shown in Fig.~\ref{figure1}(b). Under appropriate imaging conditions both Eu superstructures are resolved by STM through the covering Gr layer, as shown in Figs.~\ref{figure1}(c) and (d). Similar imaging of atomic details through Gr was  reported by Mallet \emph{et al.} \cite{Mallet2007}.

Continued Eu deposition at \unit[720]{K} does not further compress the intercalation layer beyond the $(\sqrt{3} \times \sqrt{3})\mathrm{R}30^{\circ}$ structure. Instead, the surplus Eu completely re-evaporates back into the vacuum. This fact enables a simple preparation of a perfectly filled $(\sqrt{3} \times \sqrt{3})\mathrm{R}30^{\circ}$ layer by exposing Gr/Ir(111) to a Eu excess. The saturated $(2 \times 2)$ layer can be either prepared by stepwise Eu deposition and subsequent LEED checking, or by annealing a $(\sqrt{3} \times \sqrt{3})$R$30^{\circ}$ layer to \unit[1240]{K}. After this treatment, LEED exclusively shows $(2 \times 2)$ spots and STM reveals a homogeneous intercalation layer. As the latter technique is better reproducible with less effort, we used it for the subsequent investigations.

As an alternative option to prepare an intercalated Eu layer one might consider to deposit Eu at room temperature followed by annealing to \unit[720]{K}.
While this method in fact may result in a high-quality intercalation layer,
it implies that Eu is exposed for some time to the background pressure of the UHV system prior to annealing. Due to its extremely high reactivity, non-volatile Eu compounds may form, e.g., ferromagnetic EuO by reaction with residual water. We found samples prepared by room temperature Eu deposition and subsequent annealing to be poisoned by non-reproducible spurious ferromagnetic XMCD signals, presumably resulting from trace amounts of EuO.

\section{Mechanism of intercalation}

Using DFT, we recently calculated a considerable energy gain for intercalation of Eu atoms under Gr on Ir(111) \cite{Schumacher2013a}. Nevertheless, Eu intercalation is a thermally activated process, since only adsorption, but no intercalation, takes place upon exposure of Gr/Ir(111) to Eu at \unit[300]{K} \cite{Forster2012b,Schumacher2013b}. It is therefore of interest to identify the mechanisms of intercalation and the nature of the activated process.

In the literature several mechanisms for the intercalation process are discussed: Since Eu intercalates under a fully closed Gr layer, (i) intercalation from the edges of Gr flakes towards their interior, as observed in the case of O$_2$ intercalation \cite{Granas2012, Sutter2010}, can be directly excluded here. An alternative pathway suggested for intercalation is (ii) Gr penetration at wrinkles via nano-scale cracks formed by large forces, where wrinkles of different orientation connect \cite{Petrovic2013}. Moreover, the (iii) intercalation though point-like defects in Gr has been proposed for several cases \cite{Markevich2012, Kaloni2012, Xia2012, Song2013, Duong2012}. Finally, intercalation via (iv) reactive passage by defect formation has been postulated \cite{Sicot2012, Boukhvalov2009, Virojanadara2010}.

To investigate which of the mechanisms (ii)-(iv) are of importance for Eu intercalation, the intercalation process was imaged in real time by electron microscopy. We used PEEM imaging which turned out to yield a better contrast than LEEM, probably because of its high sensitivity to the work function difference between Eu-intercalated and pristine Gr \cite{Schumacher2013b}. At an intercalation temperature of \unit[720]{K}, only a homogeneous brightness change on the whole sample is observed during Eu deposition. At this temperature Eu is either too mobile or the locations of penetration are too abundant to obtain information by PEEM. Therefore, the intercalation temperature was lowered to \unit[620]{K}. Furthermore, the temperature of the CVD step in Gr growth was increased from \unit[1170]{K} to \unit[1470]{K}. This results in more wrinkles, but less other defects, making it easier to distinguish between the relevance of both. 

\begin{figure*} [tb]
\begin{center}
\includegraphics[width=1.0\textwidth]{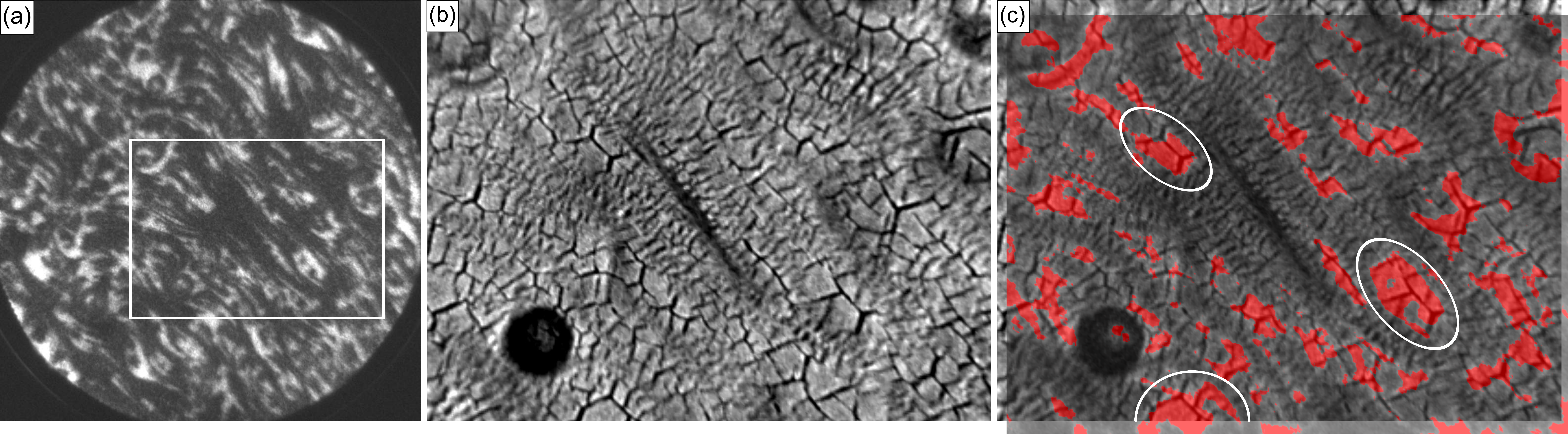}	
\caption{\label{figure2}
(color online) (a) PEEM image during Eu deposition at \unit[620]{K} ($\unit[25]{\text{\textmu}m}$ field of view). Intercalated regions are imaged bright. (b) Defocused mirror mode LEEM image ($\unit[15]{\text{\textmu}m} \times \unit[10]{\text{\textmu}m}$) of the region indicated in (a). The wrinkle network is well visible. (c) Same image as (b), but overlayed with the corresponding PEEM image. The intercalated regions colored in red are attached to the wrinkle network as especially well visible in the encircled regions.}
\end{center}
\end{figure*}  

The PEEM image in Fig.~\ref{figure2}(a) shows the early stage of intercalation. Using a PEEM series during Eu deposition (see Supplemental Material \cite{Supplement}), the bright regions are identified as the intercalated areas. Since the intercalated material is distributed inhomogeneously on a mesoscopic scale, we consider mechanism (iv), reactive passage via defect formation, to be unlikely. Figure~\ref{figure2}(b) shows a slightly defocused mirror mode LEEM image of the region highlighted by a box in Fig.~\ref{figure2}(a). Since wrinkles are substantially higher than the average surface, the slight defocus results in a pronounced contrast for the wrinkle network. When overlaying in Fig.~\ref{figure2}(c) the PEEM image of the same region (with the intercalated areas colored in red for better visibility), it becomes apparent that the intercalated material is attached to the most prominent wrinkles [compare circled locations in Fig.~\ref{figure2}(c)]. These observations indicate that mechanism (ii), penetration at wrinkles via nano-scale cracks, is operative here. 

One might argue that the Eu penetrates at random locations through mechanisms (iii) or (iv) and only accumulates at wrinkles. However, then one would expect the material to accumulate at all wrinkles, and not only the most pronounced ones, which presumably involve the largest Gr deformations at their crossing points and thus the highest probability to develop cracks.

\begin{figure} [bt]
\begin{center}
\includegraphics[width=0.7\columnwidth]{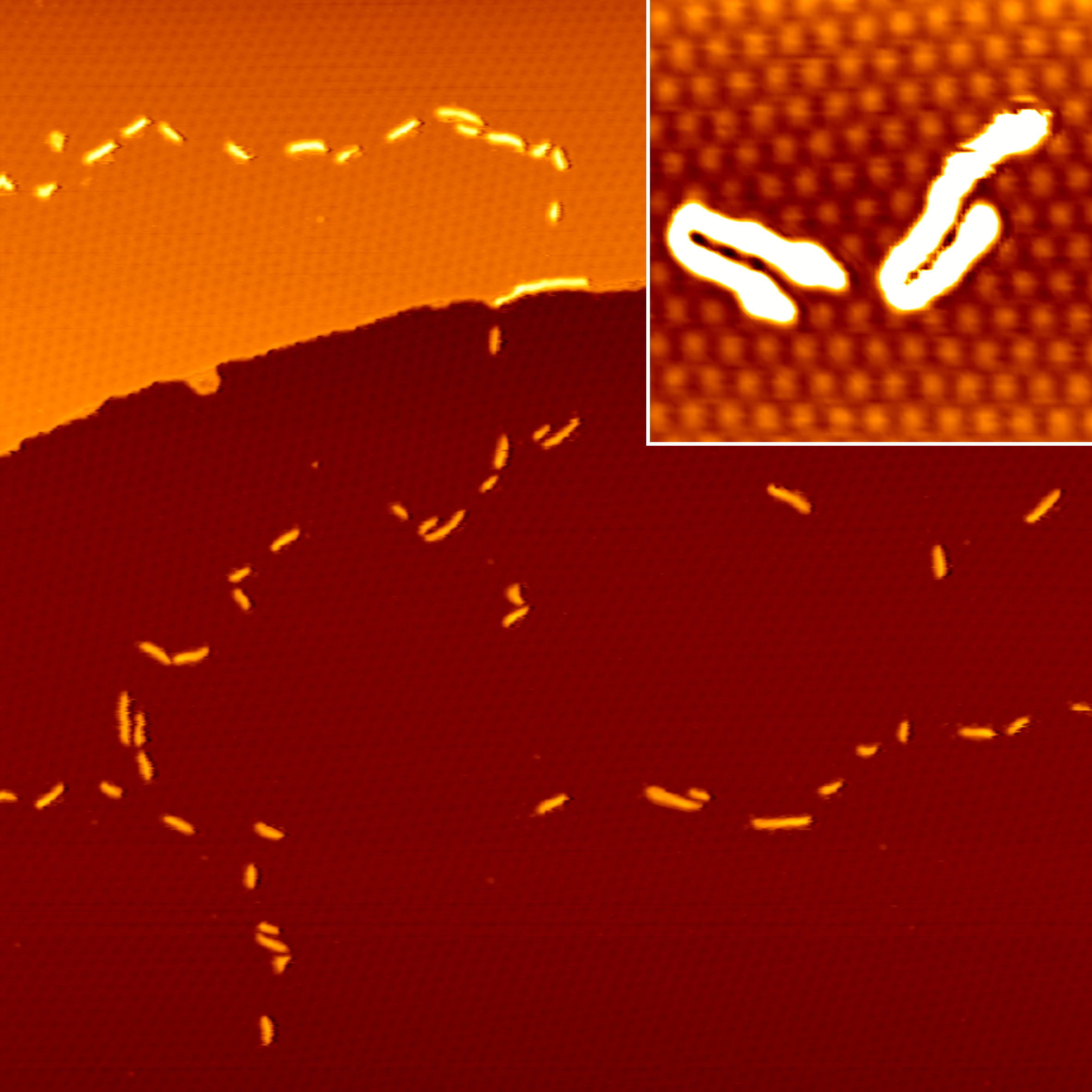}	
\caption{\label{figure3}
(color online) STM topograph ($\unit[160]{nm} \times \unit[160]{nm}$) of a small intercalated Eu amount decorating  pentagon-heptagon defects in a small-angle grain boundary in Gr. Inset: Zoom on two pentagon-heptagon defects decorated by intercalated Eu in a horseshoe shape ($\unit[30]{nm} \times \unit[30]{nm}$).}
\end{center}
\end{figure}  

Based on the present experiments, we cannot exclude that other intercalation mechanisms are operative in the absence of a wrinkle network or for different intercalation temperatures. A relevant finding in this respect is presented in Fig.~\ref{figure3}. The STM topograph displays a closed Gr layer, grown with a CVD temperature of \unit[1170]{K}, after exposure to a very small amount of Eu at \unit[720]{K}. The intercalated material is found only at pentagon-heptagon defects, which are the constitutive elements of small-angle grain boundaries in the Gr layer \cite{Coraux2008}. As shown in the inset of Fig.~\ref{figure3}, often the intercalated Eu decorates the defects in a horseshoe shape, with the open end of the horseshoe at the pentagon side of the defect. There are two possible explanations for our observation. First, pentagon-heptagon defects serve as penetration points for intercalation, either per se [mechanism (iii)] or as locations, where reactive passage becomes possible [mechanism (iv)]. Second, intercalation through wrinkle cracks in a mesoscopic distance might give rise to dilute and highly mobile Eu adatoms underneath Gr. Since pentagon-heptagon defects possess a considerable out-of-plane deformation \cite{Lehtinen2013}, they might trap Eu atoms in locations, where the bending of Gr away from the substrate creates a cavity. Consistent with the latter argument is the finding that pentagon-heptagon defects are not only the locations where intercalated material is seen for the lowest coverage, but also where even after long Eu exposure still tiny non-intercalated spots are visible. These would correspond to locations where Gr is bend towards the substrate, making it energetically rather costly to insert additional atoms in between. 

We conclude that mechanism (ii), penetration at wrinkles via nano-scale cracks, appears to be operative for the Eu deposition temperature of \unit[620]{K} and in the presence of a wrinkle network. For Gr formed by other growth methods, resulting in a differing concentration of wrinkles and point defects, as well as for other intercalation temperatures, intercalation mechanisms (iii) or (iv) cannot be excluded. Additional temperature-dependent and high-resolution imaging experiments during intercalation might resolve this issue.

\section{Magnetic measurements}

\begin{figure} [htbp]
\begin{center}
\includegraphics[width=1.0\columnwidth]{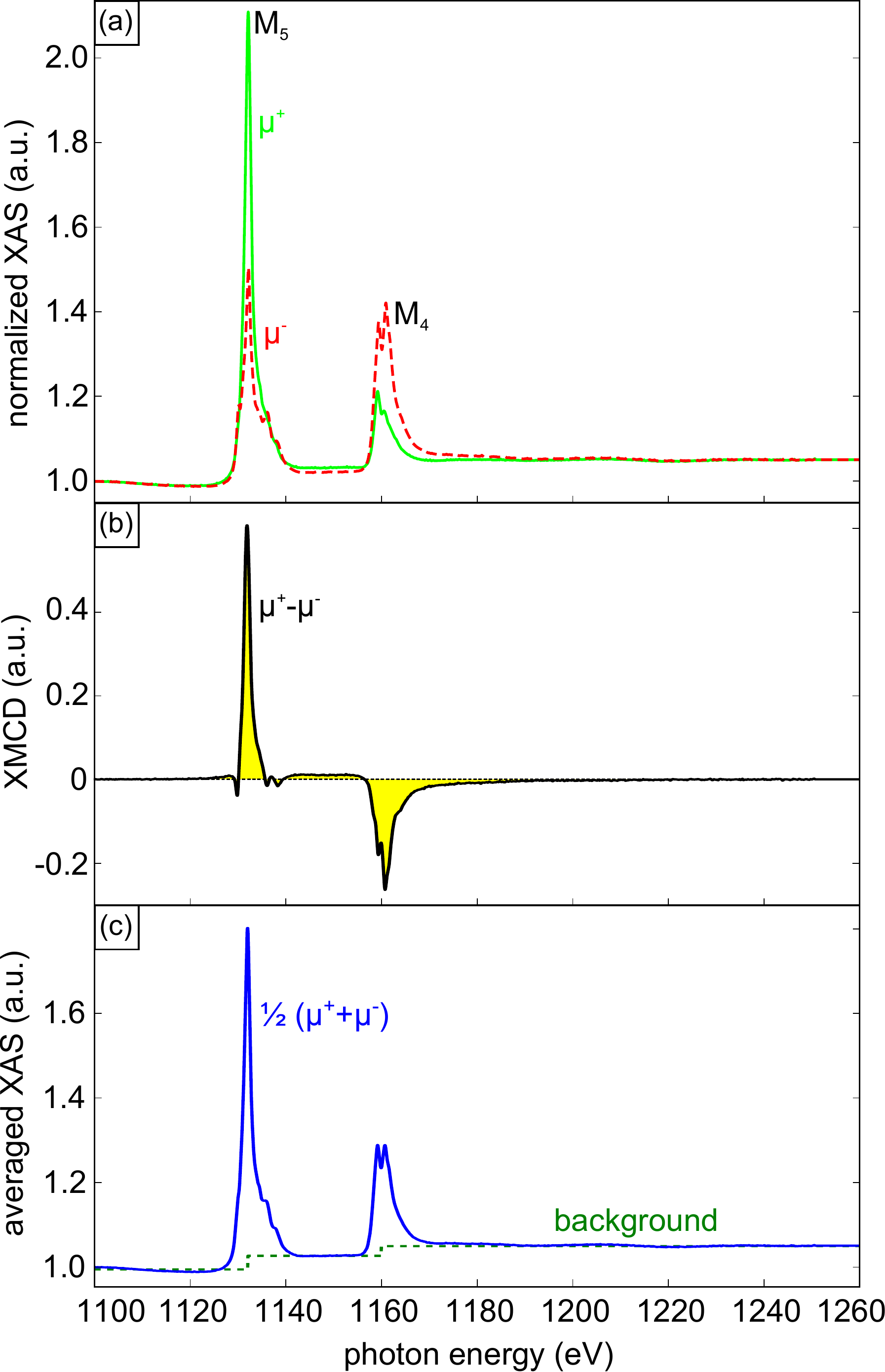}	
\caption{\label{figure4}
(color online) (a) Normalized XAS signal at \unit[10]{K} and $\unit[5]{T}$ for positive ($\mu^+$, solid green line) and negative ($\mu^-$, dashed red line) helicities in normal incidence. (b) Resulting XMCD signal ($\mu^+ - \mu^-$). (c) Polarization-averaged XAS spectrum  $\frac{1}{2}(\mu^+ + \mu^-$) (blue solid line) with step-like continuum background (green dashed line).}
\end{center}
\end{figure}

Since the magnetism of Eu originates from its 4f electrons, we investigated the Eu $M_{5,4}$ edges, i.e., transitions from 3d$_{5/2}$ and 3d$_{3/2}$ to 4f.
Figure~\ref{figure4}(a) exemplarily shows the XAS signal across those edges obtained for a fully intercalated $(\sqrt{3} \times \sqrt{3})$R$30^{\circ}$ Eu layer in normal incidence for both helicities. The external magnetic field is $B_{\mathrm{ext}}=\unit[5]{T}$ and the sample temperature $T=\unit[10]{K}$.
The spectra have been normalized to the average intensity in the pre-edge region between \unit[1100]{eV} and \unit[1120]{eV}. The corresponding spectra for the $(2 \times 2)$ intercalation structure are shown in the Supplemental Material \cite{Supplement}.

Figure~\ref{figure4}(b) shows the XMCD signal $\mu^+ - \mu^-$ which results from subtraction of the two absorption spectra. In order to deduce the orbital ($m_L$) and spin ($m_S$) magnetic moments from the XMCD data,  we used the sum rules derived by Thole \emph{et al.} \cite{Thole1992} and Carra \emph{et al.} \cite{Carra1993}. From the general formula one obtains for the M$_{5,4}$ edges 
\begin{eqnarray}
\frac{m_L}{n_{\mathrm{h}} \mu_{\mathrm{B}}} &=& \frac{q}{r} \\
\frac{m_S}{n_{\mathrm{h}} \mu_{\mathrm{B}}} &=& \frac{5p-3q}{2r} - 6 \frac{\langle T_z \rangle}{n_{\mathrm{h}}} \approx \frac{5p-3q}{2r}
\end{eqnarray}

Herein, $p$ and $q$ are the integrals of the XMCD signal over the $M_5$ and both absorption edges, respectively, and $n_{\mathrm{h}}$ denotes the number of holes in the 4f shell. The dipolar term $\langle T_z \rangle$ vanishes for zero orbital momentum~\cite{Carra1993, Wu1994, Crocombette1996}, as is the case for Eu$^{2+}$ in 4f$^7$ configuration. Evidence for the absence of other oxidation states is provided below. Therefore, we will not distinguish between real and effective spin moments in the following.

For normalization we used the integral $r$ over the polarization-averaged absorption cross section $\frac{1}{2}(\mu^+ + \mu^-$) which is shown in Fig.~\ref{figure4}(c). Schillé \emph{et al.} pointed out that if the isotropic spectrum for rare earths differs from the polarization-averaged one, it has to be taken into account, too \cite{Schille1993}. Therefore, we also recorded data at room temperature without a magnetic field applied to ensure a non-magnetic state. These spectra did not differ from the polarization-averaged ones, justifying our approach. The polarization-averaged spectrum is indistinguishable from the one for divalent Eu in Refs.~\onlinecite{Thole1985} and \onlinecite{Forster2011}, but distinct from the one of Eu$^{3+}$ in the same references. Hence, we can exclude the higher oxidation state.  

In order to separate the $M_{5,4}$ contributions from the continuum, a step-like function as depicted in Fig.~\ref{figure4}(c) has been subtracted. The heights of the plateaus were fitted in the regions before, between, and after the edges. The ratio of the step heights at the $M_5$ and $M_4$ edges (\emph{branching ratio}) agrees within 10\% with the theoretically expected value of 3:2 resulting from the degeneracy of the 3d$_{5/2}$ and 3d$_{3/2}$ orbitals. 

\begin{table}[htbp]
\caption{Spin moment $m_S$ per Eu atom derived from the sum rules (for $n_{\mathrm{h}}=7$), and zero-field susceptibilities for intercalated layers of different density. The data were taken at \unit[10]{K} under a field of \unit[5]{T}. X-rays and magnetic field were both incident normally ($0^{\circ}$) or grazing ($60^{\circ}$). We estimate the relative errors to be 10\%.}
\begin{ruledtabular}
\begin{tabular}{lcccccc}
 & & \multicolumn{2}{c}{$m_S (\unit{\mu_{\mathrm{B}}})$}  & & \multicolumn{2}{c}{$\chi (\mathrm{\mu_{\mathrm{B}}/(T\cdot\mathrm{atom})})$} \\ 
sample & & $0^{\circ}$ & $60^{\circ}$ & & $0^{\circ}$ & $60^{\circ}$ \\ 
$(2 \times 2)$ & & 4.9 & 5.3 & & 1.2 & 1.5 \\ 
$(\sqrt{3} \times \sqrt{3})$R$30^{\circ}$ & & 6.3 & 6.8 & & 6.2 & 24 \\ 
\end{tabular}
\end{ruledtabular}

\label{tab:momentsEuintercalation}
\end{table}

By applying the orbital sum rule to the XMCD data measured at \unit[10]{K} and \unit[5]{T} for incidence angles of  $0^{\circ}$ and $60^{\circ}$ we find $m_L$ to be zero within the error of our measurement for both structures. This implies that Eu is present in the half-filled 4f$^7$ configuration, in agreement with our assumptions above. Thus, we specify the spin moments listed in Tab.~\ref{tab:momentsEuintercalation} per Eu atom using $n_{\mathrm{h}}=7$. We also measured data for $20^{\circ}$ and  $40^{\circ}$ incidence angle yielding consistent results.

\begin{figure} [b]
\begin{center}
\includegraphics[width=1.0\columnwidth]{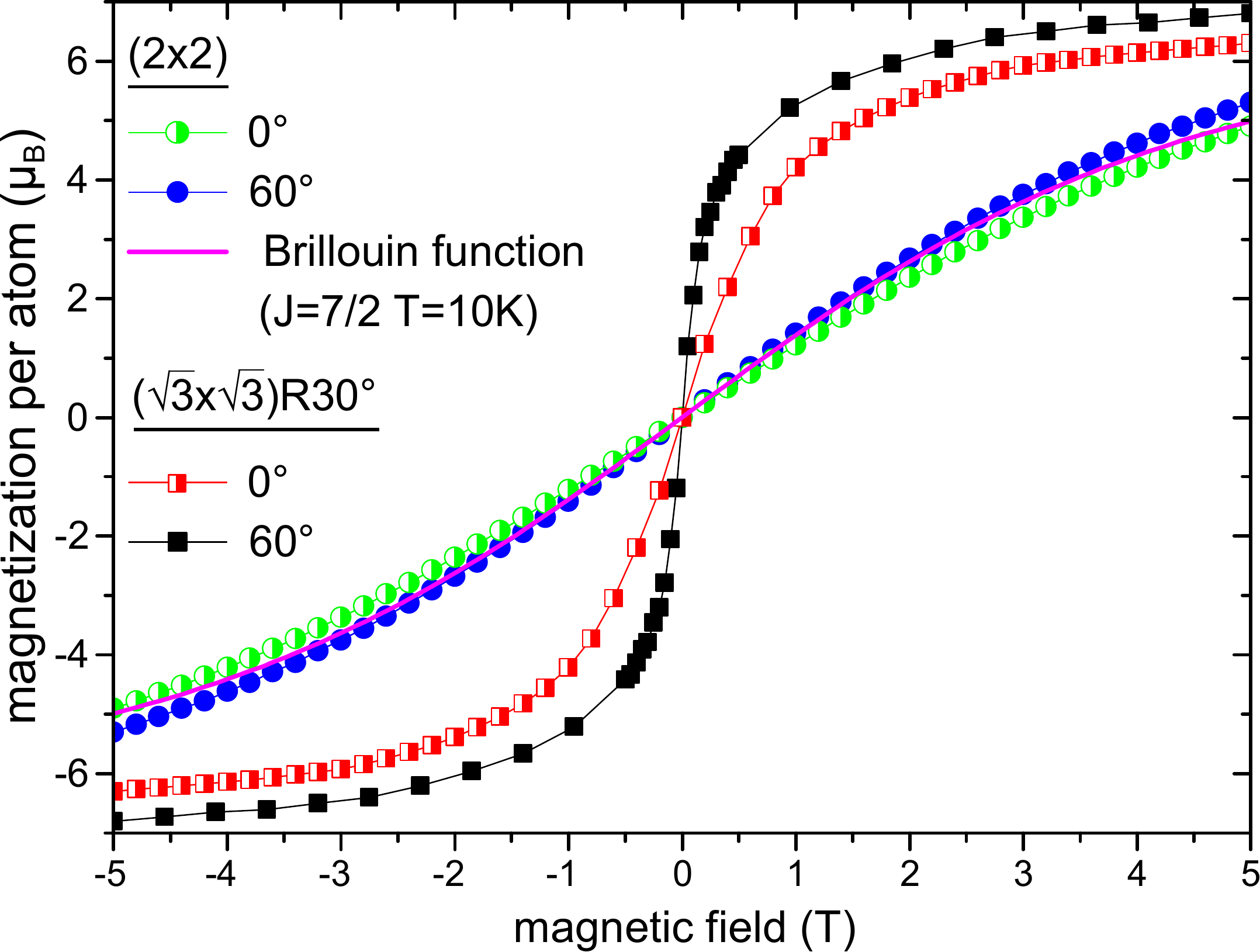}	
\caption{\label{figure5}
(color online) Magnetization loops for a fully intercalated $(\sqrt{3} \times \sqrt{3})\mathrm{R}30^{\circ}$ Eu layer (squares) and a $(2 \times 2)$ Eu layer (circles) at \unit[10]{K} for normal (half filled) and grazing (solid) incidence. Each curve is scaled to the corresponding effective spin moment per Eu atom at \unit[5]{T}.}
\end{center}
\end{figure}  

To learn more about the magnetic behavior, we measured magnetization \textit{versus} field loops, which were obtained by normalizing the XAS signal obtained at the Eu M$_5$ edge to a pre-edge value in dependence on the magnetic field. The loops for normal and grazing incidence are shown in Fig.~\ref{figure5} for a fully intercalated $(\sqrt{3} \times \sqrt{3})$R$30^{\circ}$ Eu layer (squares) and a saturated $(2 \times 2)$ Eu layer (circles). As we do not observe any hysteresis, the loops were averaged over increasing and decreasing field direction. Furthermore, the data were symmetrized, i.e., the magnetization $M(B_{\mathrm{ext}})$ was replaced by $\frac{1}{2}[M(B_{\mathrm{ext}})-M(-B_{\mathrm{ext}})]$ as we expect $M(-B_{\mathrm{ext}})=-M(B_{\mathrm{ext}})$. Each curve is scaled using the corresponding spin moment per Eu atom at \unit[5]{T} from Tab.~\ref{tab:momentsEuintercalation}. As a characteristic quantity we determine the zero-field susceptibility by fitting the curve linearly at $B_{\mathrm{ext}}=0$. We note that this method probably underestimates the susceptibility due to the limited resolution. Therefore, the values listed in Tab.~\ref{tab:momentsEuintercalation} should be interpreted as lower limits.

\section{Discussion of the magnetic properties}

First, we address the measurements concerning the intercalated $(2 \times 2)$ structure. The magnetization curves for normal and grazing incidence are nearly identical, show an almost linear behavior with a low zero-field susceptibility of about $\unit[1.4]{\mu_B/(T\cdot\mathrm{atom})}$, and do not saturate at \unit[5]{T}. The slightly flatter magnetization curve for normal compared to grazing magnetic field points to a weak easy-plane anisotropy. The flat and almost isotropic magnetization behavior is indicative for a paramagnetic system. To confirm this hypothesis, we recall that the field-dependent magnetization $M(B)$  per atom for a paramagnet at a temperature $T$ is given by the Brillouin function 

\begin{small}
\begin{equation}
\label{eq:langevin}
M(B)= g \mu_{\mathrm B} J \left( \frac{2J+1}{2J} \coth \left( \frac{2J+1}{2J} x \right) - \frac{1}{2J} \coth \left( \frac{1}{2J} x \right) \right)
\end{equation}
\end{small}

with $x=\frac{g \mu_{\mathrm B} J B}{k_{\mathrm{B}} T}$. $J$ denotes the total angular momentum ($J=\frac{7}{2}$ for $S=\frac{7}{2}$ and $L=0$) and g is the Landé factor ($g=2$ for Eu). The function $M(B)$ is plotted in Fig.~\ref{figure5} as magenta line. Comparing the curve to the measured data yields an excellent agreement except for the weak easy-plane anisotropy to be discussed later. Therefore, we conclude that the intercalated $(2 \times 2)$ structure behaves like an ordinary paramagnet. 

In contrast to the $(2\times 2)$ intercalation structure, in the $(\sqrt{3} \times \sqrt{3})$R$30^{\circ}$ layer the moments are almost saturated at \unit[5]{T}. The saturation value is close to the expected one of $\unit[7]{\mu_{\mathrm{B}}}$ for the 4f$^7$ configuration. The measurements further show a much higher zero-field susceptibility of $\unit[6.2]{\mu_B/(T\cdot\mathrm{atom})}$ in normal and $\unit[24]{\mu_B/(T\cdot\mathrm{atom})}$ in grazing incidence compared to the paramagnetic $(2 \times 2)$ layer. These observations give strong evidence for a significant ferromagnetic coupling. 

To explain the fundamentally different magnetic behavior of the two structures, we consider that magnetic coupling between Eu moments is mediated by the RKKY interaction, as for the first-stage Eu GIC EuC$_6$. The RKKY interaction has an oscillatory nature and can be ferro- or antiferromagnetic depending on the distance.  Indeed, for EuC$_6$ the transition between ferro- and antiferromagnetic RKKY coupling occurs at around the interatomic distance present in the $(2 \times 2)$ structure \cite{Chen1986}, thereby inhibiting magnetic coupling.

\begin{figure} [htbp]
\begin{center}
\includegraphics[width=1.0\columnwidth]{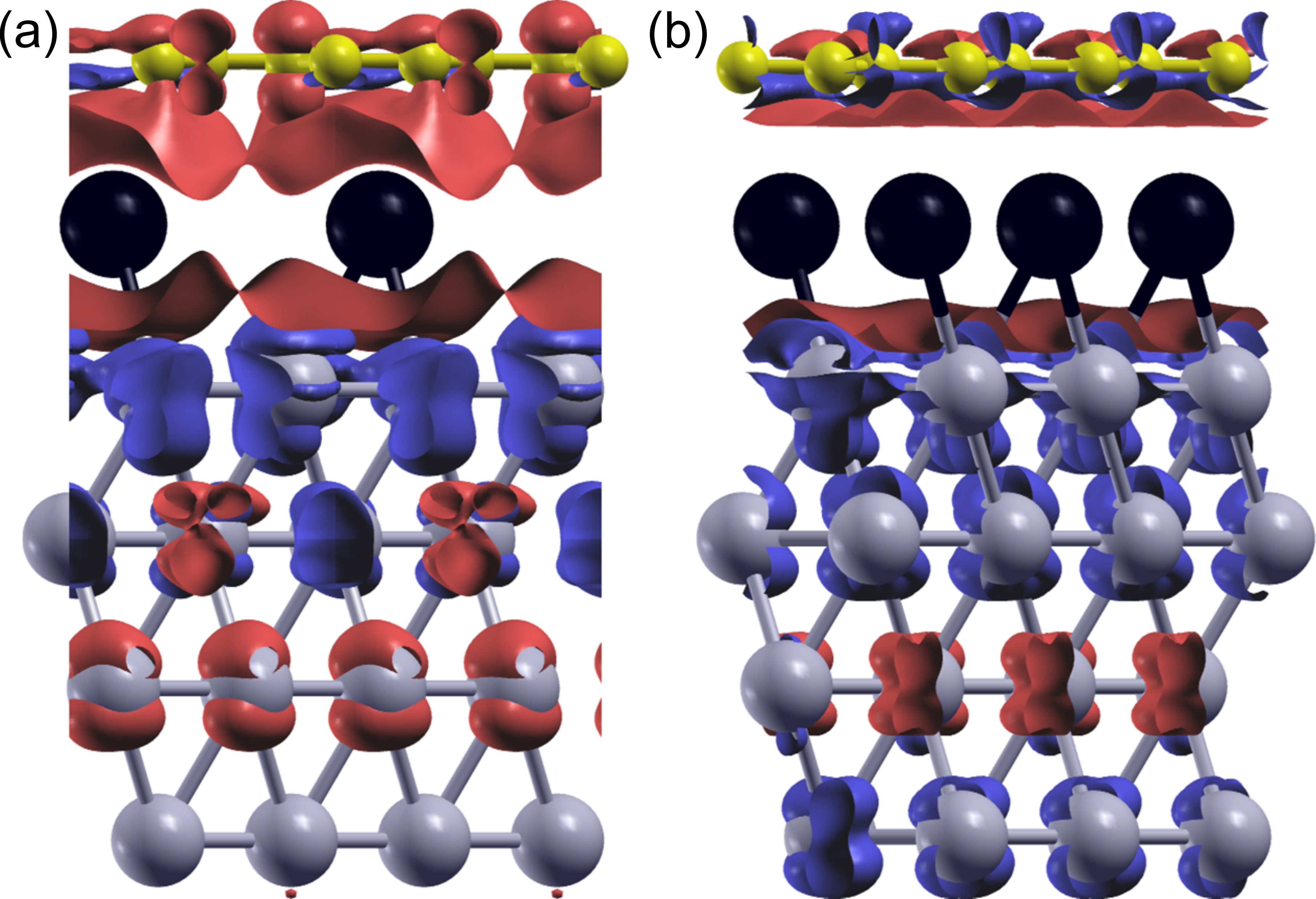}	
\caption{\label{figure6}
(color online) Spin densities for Eu (black) intercalated between Gr (yellow) and Ir(111) (gray) for (a) $(2 \times 2)$, and (b) $(\sqrt{3} \times \sqrt{3})$R$30^{\circ}$ structures. Isosurfaces of the spin density ($|m(r)|=\unit[0.001]{e/\text{\AA}^3}$) are shown. The majority spin isosurface ($m>0$) is shown in red, the minority one ($m<0$) in blue.}
\end{center}
\end{figure}  

However, in the present case the Ir substrate has to be taken into account. To this end, we calculated the spin densities by DFT, which are shown in Fig.~\ref{figure6} for the $(2 \times 2)$ and $(\sqrt{3} \times \sqrt{3})$R$30^{\circ}$ intercalation structures. In both cases, a substantial polarization of the Ir substrate is evident, which largely exceeds the one of the Gr layer. For the $(2 \times 2)$ structure, we calculate a magnetic moment per C atom of $\unit[0.001]{\mu_B}$, whereas we get $\unit[-0.0095]{\mu_B}$ in the Ir surface layer, which is larger by one order of magnitude. For the  $(\sqrt{3} \times \sqrt{3})$R$30^{\circ}$ structure, the magnetic moment per C atom of $\unit[-0.001]{\mu_B}$ is again small, but with reversed sign. Also here, the moment per Ir surface atom of $\unit[-0.017]{\mu_B}$ is substantially larger than the C magnetization, and also larger than the Ir surface magnetization for the $(2 \times 2)$ structure. The main difference between the two structures occurs in the Ir subsurface layer: Whereas the $(2 \times 2)$ structure induces alternating Ir moments that average to a small value of $\unit[-0.003]{\mu_B}$ per Ir surface unit cell area, for the $(\sqrt{3} \times \sqrt{3})$R$30^{\circ}$ structure a homogeneous spin polarization of $\unit[-0.012]{\mu_B}$ per Ir surface unit cell is found.  

Based on the calculations we tentatively conclude: (1)~The RKKY interaction through the Ir substrate, rather than through Gr, is dominating the magnetic Eu-Eu interaction, because the induced magnetic moments are larger for Ir than for the C layer. (2) The coupling for the  $(\sqrt{3} \times \sqrt{3})$R$30^{\circ}$ structure is stronger than for the $(2 \times 2)$ structure, because the magnetic moments in the first and second Ir surface layer are larger. 

Although we find evidence for ferromagnetic coupling in the  $(\sqrt{3} \times \sqrt{3})$R$30^{\circ}$ structure, there is no spontaneous magnetization or hysteresis. 
This indicates that at 10~K the layer is either (i) just above its Curie temperature $T_{\mathrm{C}}$ or (ii) composed of magnetic units that display superparamagnetic behavior. Finally, one might argue that (iii) according to the Mermin-Wagner theorem in two dimensions there cannot be spontaneous magnetization at all at non-zero temperature \cite{Mermin1966}.

Considering (i), we note that according to the Curie-Weiss law the susceptibility of a ferromagnet diverges as $T$ approaches $T_{\mathrm{C}}$ from above, which explains the high susceptibility without hysteresis if measuring close to, but above $T_{\mathrm{C}}$. Based on the calculated Curie constant $C$ for $J=\frac{7}{2}$, we estimate that our measurement temperature would need to be just a few percent above $T_{\mathrm{C}}$ to get the observed susceptibility $\chi$.

%
%

Considering (ii), the superparamagnetic unit could be constituted by the moiré unit cell given that the Eu atoms' adsorption positions relative to the Ir surface atoms (e.g., top, hollow, bridge) are dependent on the position within the moiré unit cell and so will the magnetic coupling between these Eu atoms. Indeed, calculating the paramagnetic susceptibility of a magnetic moment equal to the sum of Eu moments in a moiré unit cell at our measurement temperature yields $\unit[37]{\mu_B/(T\cdot\mathrm{atom})}$, which is not far from the $\unit[24]{\mu_B/(T\cdot\mathrm{atom})}$ experimentally determined for grazing incidence, where the demagnetization effect (see below) is smallest.

Finally, considering (iii), we note that the Mermin-Wagner theorem is based on an \emph{isotropic} model. As we will argue below, magnetocrystalline anisotropy is indeed absent or at least negligible in our system.
However, it has been shown that the long-range dipolar interaction present in our case is sufficient to change the magnon dispersion such that the Mermin-Wagner theorem no longer applies \cite{Grechnev2005}.
In this case, there is a temperature $T_1 \propto JS^2$ at which short-range order occurs, and a logarithmically lower temperature $T_{\mathrm{C}} \propto JS^2/\ln(J/\Delta)$ below which long-range order evolves. Herein, $\Delta$ is an energy determined by the dipole-dipole interaction. The logarithmic factor is typically on the order of 20 \cite{Grechnev2005}. It may be plausible that our measurement temperature lies between these two temperatures, where one would expect long-wavelength spin waves to lead to an average magnetization of zero in the zero-field case, but a much larger response of the magnetization to small fields than in the paramagnetic case. Unfortunately, a simple analytical solution to this problem does not exist.

In conclusion, it is understood that the $(\sqrt{3} \times \sqrt{3})$R$30^{\circ}$ structure's large susceptibility must arise from ferromagnetic regions, although it remains unclear whether these regions are limited in size by thermal fluctuations, as in (i) and (iii), or are of structural origin, as in (ii). Temperature-dependent magnetization loop measurements should allow to resolve this issue.

\section{Origin of the magnetic anisotropy}


The two intercalation structures both display an easy-plane anistropy, weak in the case of the $(2 \times 2)$ and rather strong in the case of the $(\sqrt{3} \times \sqrt{3})$R$30^{\circ}$ structure. We consider two sources of anisotropy: (i) \emph{magnetocrystalline} and (ii) \emph{shape} anisotropy. 

(i) As we find $m_L=0$ and thus $\vec{L}=0$, we can exclude the presence of magnetocrystalline anisotropy in the Eu layer, which is caused by spin-orbit coupling and thus proportional to $\vec{L} \cdot \vec{S}$. For the contributions from induced moments in the Ir atoms, where $m_L\neq 0$, an upper bound can be estimated which is well below what we find for the shape anisotropy energy.
Therefore, we discard magnetocrystalline anisotropy as insignificant.

(ii) In contrast, the shape anisotropy energy per atom in our system is expected to be large, because the layer is ultimately thin at only one monolayer, and the magnetic moment of Eu of $\unit[7]{\mu_\mathrm{B}/\mathrm{atom}}$, which enters quadratically, is large as well. Using a simple continuum treatment (cf.\ Eq.~(3) in Ref.~\onlinecite{Johnson1996}), and assuming 
a layer thickness equal to the Eu-Eu distance in the respective structure,
we obtain shape anisotropy energies per atom of \unit[160]{\text{\textmu}eV} [for $(2 \times 2)$] and \unit[246]{\text{\textmu}eV} [for $(\sqrt{3} \times \sqrt{3})$R$30^{\circ}$].
For comparison, these values are large, in the range of what can be found in other systems only in the form of \emph{magnetocrystalline} anisotropy energy (e.g., $\unit[60]{\text{\textmu} eV}$ for hcp Co \cite{Daalderop1990} or $\unit[3000]{\text{\textmu} eV}$ for L1$_0$ FePt \cite{Barmak2005}). 

We will consider shape anisotropy as a demagnetization effect.
Given a volume susceptibility $\chi_\mathrm{v}$, the apparent susceptibility will be
\begin{equation}\label{eq:demagfac}
\chi_\mathrm{v, app} = \frac{\chi_\mathrm{v}}{1+N\chi_\mathrm{v}}
\end{equation}
when the demagnetization factor $N$ is taken into account.
For an external field in the plane of an infinitely extended layer it is $N=0$, and perpendicular to the plane it is $N=1$.
Under an angle $\theta$ to the surface normal, a mixing according to
\begin{equation}\label{eq:demagfac2}
\chi_\mathrm{v, app} = \frac{\chi_\mathrm{v}}{1+\chi_\mathrm{v}}\cos^2\theta + \chi_\mathrm{v} \sin^2\theta
\end{equation}
can be derived by appropriately treating the demagnetization factor as a tensor.
The volume susceptibility is calculated from the $\chi$ used above, which was in units of $\mu_\mathrm{B} / (\mathrm{T} \cdot \mathrm{atom})$, according to
$\chi_\mathrm{v} = \frac{\mu_0}{V} \cdot \chi$
with $V$ the volume occupied by one Eu atom, again
assuming a layer thickness equal to the Eu-Eu distance.

For the $(2 \times 2)$ structure we use the theoretically expected susceptibility of a paramagnetic $J=7/2$ ion at $T=\unit[10]{K}$ to obtain $\chi(0^\circ)=\unit[1.22]{\mu_\mathrm{B}/(T\cdot\mathrm{atom})}$ and $\chi(60^\circ)=\unit[1.36]{\mu_\mathrm{B}/(T\cdot\mathrm{atom})}$, in good agreement with the experimental data given that 10\% is already our estimate of the error in $\chi$.

Now concerning the $(\sqrt{3} \times \sqrt{3})$R$30^{\circ}$ structure, where we \emph{almost} have a ferromagnet, the stronger anisotropy compared to the $(2 \times 2)$ structure can as well be understood with Eq.~\ref{eq:demagfac2} within better than a factor of two by taking into account the larger susceptibility.



As pointed out by Johnson \emph{et al.} \cite{Johnson1996}, for a magnetic monolayer, calculations based on discrete dipoles should be preferred over a continuum treatment as performed above. We therefore conducted such discrete calculations (see Supplemental Material \cite{Supplement}) and found only slight quantitative deviations for the anisotropy energy and zero-field susceptibility from our continuum treatment.

In conclusion, the observed anisotropy is understood for both intercalation structures to result predominantly from shape, while the magnetocrystalline contribution appears to be negligible.

\newcommand{\EuC}{EuC$_6$}
\section{Comparison with the graphite intercalation compound \texorpdfstring{\protect\EuC}{EuC6}}
\begin{figure}[b]
\begin{center}
\includegraphics[width=1.0\columnwidth]{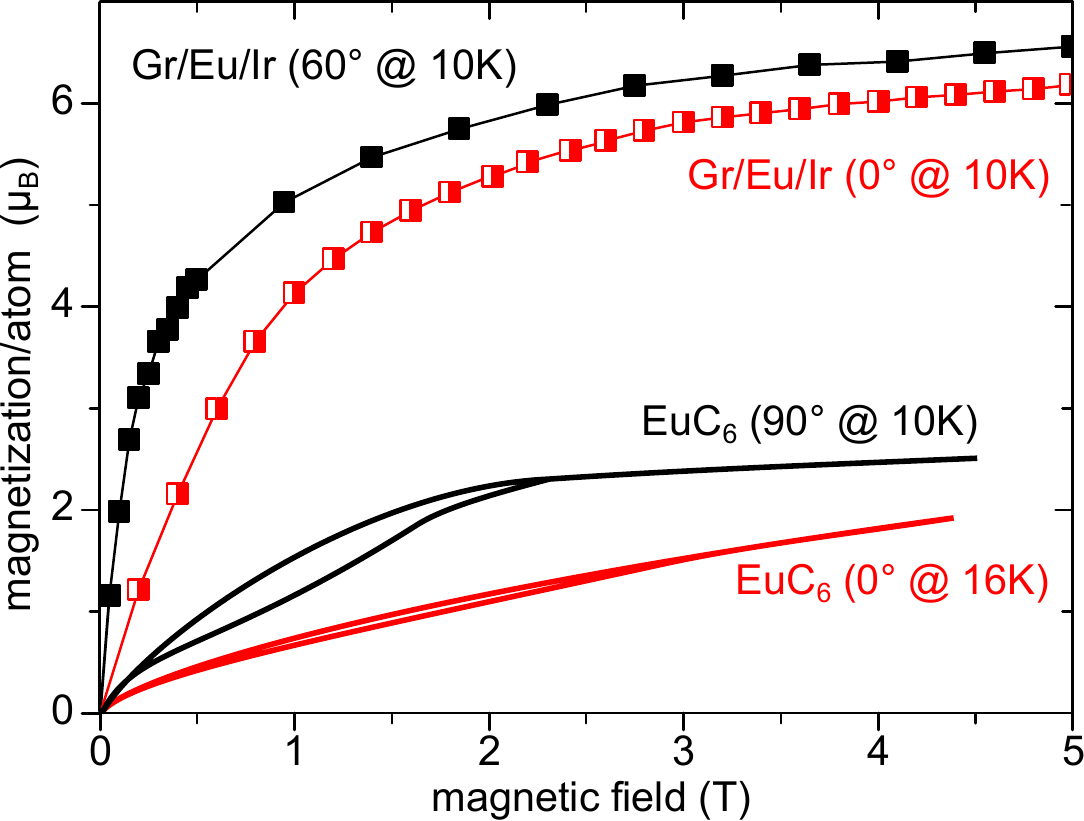}	
\caption{\label{figure7} 
(color online) Magnetization loops for a fully intercalated $(\sqrt{3} \times \sqrt{3})$R$30^{\circ}$ Eu layer at the indicated temperatures and incidence angles (same data as in Fig.~\ref{figure5}). Additionally, corresponding data for EuC$_6$ taken from Ref.~\onlinecite{Suematsu1981a} are shown.}

\end{center}
\end{figure}
It is tempting to compare the magnetic properties of the $(\sqrt{3} \times \sqrt{3})$R$30^{\circ}$ intercalation structure with those of the first-stage Eu GIC EuC$_6$ as both have the same in-plane  structure. As already stated in the introduction, EuC$_6$ shows metamagnetic behavior \cite{Suematsu1983, Chen1986}: In the low-field range neighboring spins include angles of $120^{\circ}$, which is a typical ground state for a frustrated antiferromagnet on a triangular lattice \cite{Deutscher1993}. Upon applying higher fields the $120^{\circ}$ order changes to ferrimagnetic order, i.e., two of three spins align parallel and one antiparallel to the magnetic field, yielding a plateau at one third of the saturation magnetization. By further increasing the field, the spin structure changes to a canted spin state and finally turns to be ferromagnetic at fields higher than \unit[20]{T}. Above a temperature of \unit[40]{K} the system stays paramagnetic for all fields. 

In Fig.~\ref{figure7} the first quadrant of our magnetization loops from Fig.~\ref{figure5} is shown together with data for EuC$_6$ adapted from Ref.~\onlinecite{Suematsu1981a}. Unfortunately, no directly comparable data are available as we can only measure under a grazing angle of $60^{\circ}$, while bulk measurements allow fields exactly in-plane. Furthermore, for normal field data are available only at \unit[16]{K} instead of \unit[10]{K}.

Both systems show a pronounced easy-plane anisotropy, but beside this similarity the magnetic behavior is very different: While for the intercalation layer the initial magnetization steeply rises and goes into saturation close to $\unit[7]{\mu_B}$, the magnetization curve of EuC$_6$ has a much lower zero-field susceptibility and already saturates at about $\unit[2.5]{\mu_B}$. The low zero-field susceptibility is due to an antiferromagnetic $120^{\circ}$ spin order which changes to ferrimagnetic order yielding a saturation at $\unit[2.5]{\mu_B}$. 

We conjecture two reasons for the differing magnetic behavior: First, the Eu layer is in contact to the Ir substrate on one side if intercalated underneath Gr, while it is embedded between two Gr layers in the GIC. This is expected to substantially change the RKKY coupling, since we have shown before that the Ir substrate predominantly contributes to the RKKY interaction. Second, it has been shown that a small coupling between adjacent Eu layers is an important ingredient to explain the metamagnetic behavior of EuC$_6$ \cite{Chen1986}. This coupling is of course absent in our system, as it is restricted to a single Eu layer.

\section{Band structure and doping}

\begin{figure} [htbp]
\begin{center}
\includegraphics[width=1.0\columnwidth]{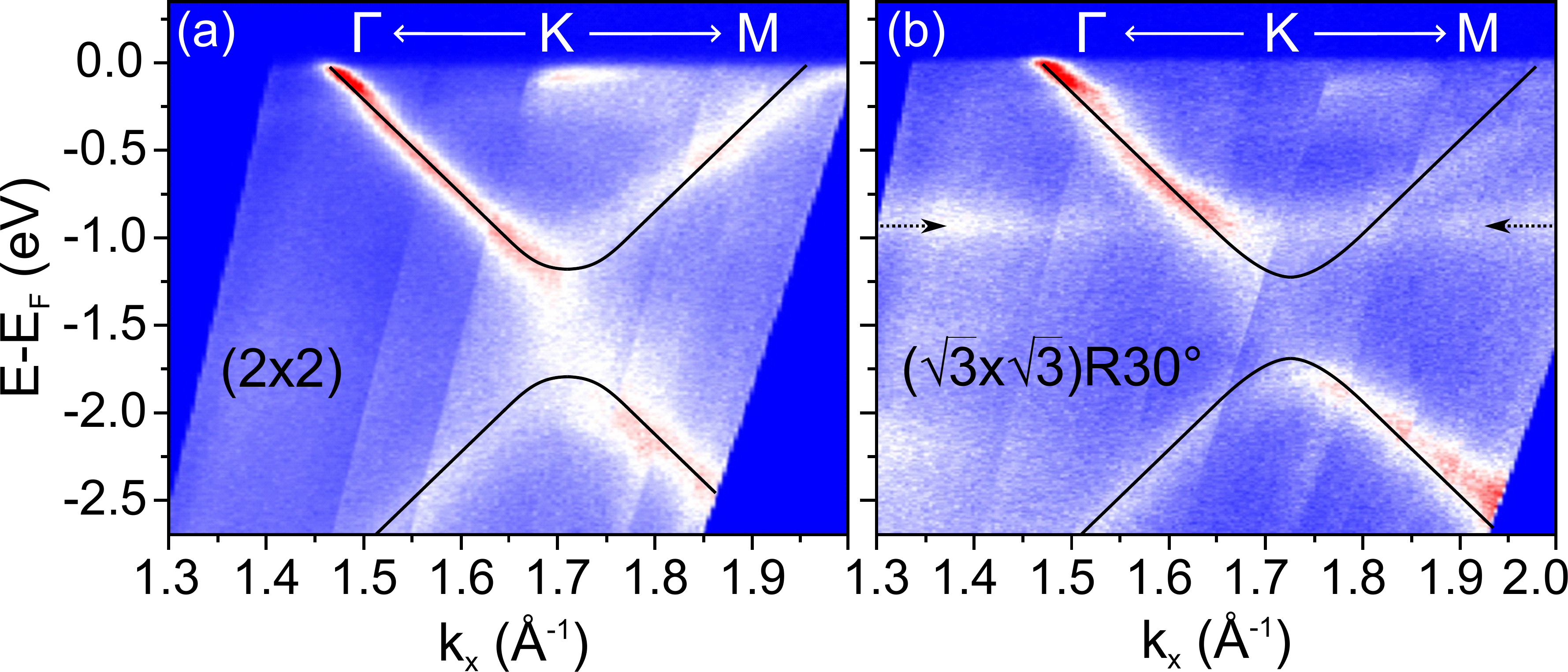}	
\caption{\label{figure8} 
(color online) ARPES spectra of Eu intercalated Gr on Ir(111) in $\Gamma$KM direction, i.e. along $k_x$. An azimuthal offset $\Delta \phi$ mimics a band gap at the Dirac cone. Thin black lines indicate fitted tight-binding bands in the nearest-neighbor approximation including the azimuthal offset as fit parameter \cite{CastroNeto2009,Kralj2011}.
(a) $(2 \times 2)$ structure with an azimuthal offset $\Delta \phi \approx 1.5^{\circ}$. (b) $(\sqrt{3} \times \sqrt{3})$R$30^{\circ}$ structure with $\Delta \phi \approx 1^{\circ}$. The arrows in (b) highlight a flat band, see text.}
\end{center}
\end{figure}

To investigate the electronic structure of Gr/Eu/Ir(111), we performed ARPES measurements. Figure~\ref{figure8} shows ARPES spectra in the vicinity of the K point taken along the $\Gamma$KM direction for a $(2 \times 2)$ and a $(\sqrt{3} \times \sqrt{3})$R$30^{\circ}$ intercalation layer, respectively.
Both show a graphene band structure shifted down in energy by about \unit[1.4]{eV} due to strong n-doping. The noticeable intensity anisotropy between the two branches is due to the electronic chirality of Gr \cite{Mucha2008}.
Due to the moiré superperiodicity, minigaps are present for Gr/Ir(111) in the $\pi$ band about $\unit[-0.5]{eV}$ to $\unit[-1]{eV}$ below the Fermi level \cite{Pletikosic2009}.
For Gr/Eu/Ir(111) such minigaps are absent, indicating a substantially reduced superperiodic potential.
Due to a small azimuthal misalignment with respect to the $\Gamma$KM direction, the precise location of the K point is missed in Fig.~\ref{figure8} and the cone sections of the $\pi$- and $\pi^{\ast}$-band seemingly form a band gap of about \unit[500]{meV} (see Supplemental Material for geometry \cite{Supplement}).

\begin{figure*} [htbp]
\begin{center}
\includegraphics[width=0.7\textwidth]{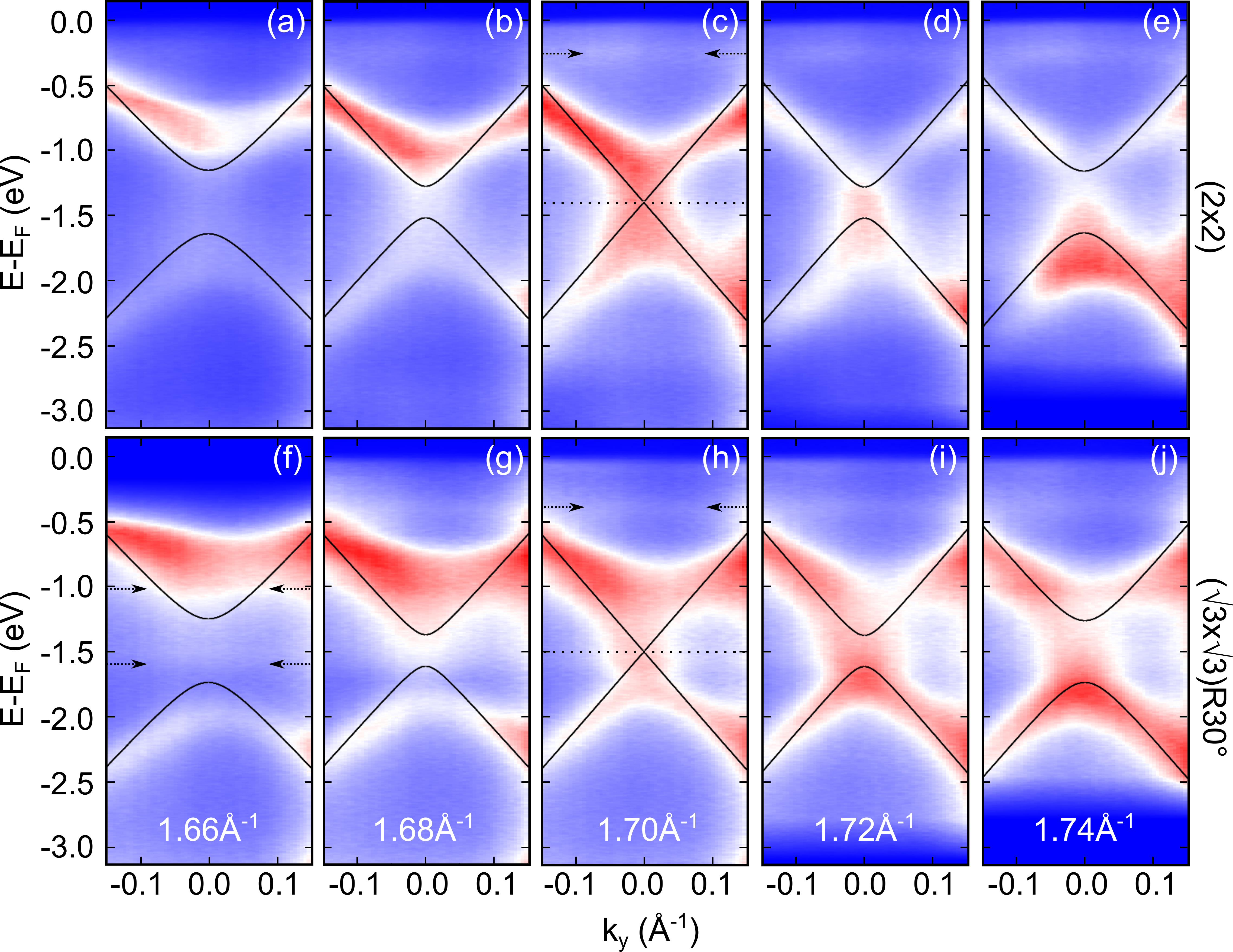}	
\caption{\label{figure9} 
(color online) ARPES spectra of Gr intercalated by Eu in (a)\,-\,(e) $(2 \times 2)$  and (f)\,-\,(j) $(\sqrt{3} \times \sqrt{3})$R$30^{\circ}$ structure.
All spectra are taken perpendicular to the $\Gamma$KM direction at the indicated values of $k_x$ on this line.
The black lines correspond to the $\pi/\pi^*$ bands in TBA rigidly shifted by $E_{\mathrm{D}}=\unit[-1.36]{eV}$ and $E_{\mathrm{D}}=\unit[-1.43]{eV}$, respectively.
The arrows highlight flat bands, see text.}
\end{center}
\end{figure*}

To investigate the electronic structure around the K point in detail, we took maps perpendicular to the $\Gamma$KM direction (i.e., along $k_y$) for fixed values of $k_x$ on the $\Gamma$KM line (see Supplemental Material for geometry \cite{Supplement}).
The data obtained for the $(2 \times 2)$ intercalation structure are shown in Figs.~\ref{figure9}(a)\,-\,(e). We compare the band structure to the well-known result from the tight-binding approximation (TBA) \cite{CastroNeto2009} with a nearest-neighbor hopping energy of $\unit[-2.848]{eV}$ taken from Ref.~\onlinecite{Kralj2011}. The doping level $E_{\mathrm{D}}$ was fitted to the data. At the K point in Fig.~\ref{figure9}(c), we find best coincidence for a doping level of $E_{\mathrm{D}} = \unit[-1.36]{eV}$.

Figure~\ref{figure9}(c) displays a perpendicular scan precisely through the K point. When moving away from the K point towards the $\Gamma$ point in Figs.~\ref{figure9}(b) and (a) the measured bands quickly deviate from the TBA band structure. Such deviations are not unexpected, as the TBA does not include many-body interactions like electron-phonon and electron-electron interactions or plasmon excitations \cite{Bostwick2007, Tse2007}.
Moving away from the K point towards the M point in Figs.~\ref{figure9}(d) and (e), the opening of the measured bands is much better reproduced by the TBA.

In addition to the Dirac cone a non-dispersing band is visible at about $\unit[-0.25]{eV}$, as highlighted by two arrows in Fig.~\ref{figure9}(c). We assign this band to the Ir surface state $S_1$ (compare Ref.~\onlinecite{Pletikosic2009} for nomenclature), which is close to the Fermi edge for bare Gr on Ir(111), while here it is shifted down by doping from the intercalated Eu layer. 

The ARPES data for the $(\sqrt{3} \times \sqrt{3})$R$30^{\circ}$ structure are shown in Figs.~\ref{figure9}(f)\,-\,(j). At the K point in Fig.~\ref{figure9}(h) the TBA fits to the measured cone using $E_{\mathrm{D}}=\unit[-1.43]{eV}$, which corresponds to a slightly higher doping level compared to the $(2 \times 2)$ intercalation structure.
Moving away from the K point, the measured data are again better reproduced by TBA in KM direction than in K$\Gamma$ direction. The flat band, which we have related before to the Ir surface state $S_1$, is still present, but now shifted to a lower energy of $\unit[-0.4]{eV}$. This also indicates a higher doping of Ir by the $(\sqrt{3} \times \sqrt{3})$R$30^{\circ}$ compared to the $(2 \times 2)$ intercalation layer.

The $\pi^*$ band is broader for the $(\sqrt{3} \times \sqrt{3})$R$30^{\circ}$ intercalation structure than for the $(2 \times 2)$ layer.
This might be related to hybridization with a flat band at $\unit[-1.0]{eV}$, which is highlighted in Fig.~\ref{figure8}(b) by arrows and is faintly visible in Fig.~\ref{figure9}(f) as well.
In principle, the surface state $S_2$ of pristine Ir(111) is located at this energy (compare Ref.~\onlinecite{Pletikosic2009}), but this state should then also be present for the $(2 \times 2)$ intercalation structure.
As we do not observe it there, the band is probably related to Eu. Finally, we note that there seems to be another non-dispersing band around $\unit[-1.6]{eV}$ as indicated in Fig.~\ref{figure9}(f), but it is very faint and of unknown origin. 

For Gr aligned with its dense packed rows to the Ir(111) substrate, as used in our experiments, Starodub \emph{et al.} \cite{Starodub2011} found the characteristic G and G' (or 2D) peaks in Raman spectroscopy to be absent.
The quenching of these Raman active Gr phonons was interpreted to result from the hybridization of Gr with the flat Ir $S_1$ surface state located directly at the Fermi edge.
Gr phonons then cause excitations of electrons in the $S_1$ state with high efficiency, thereby decreasing the phonon lifetime to an extent leaving them undetectable in Raman spectroscopy.

Here, we find that upon formation of the $(\sqrt{3} \times \sqrt{3})$R$30^{\circ}$ Eu intercalation structure, the surface state $S_1$ is downshifted to $\unit[-0.4]{eV}$ [compare Fig.~\ref{figure9}(g)].
Since the energy of the Gr phonons is below \unit[0.2]{eV}, excitation of electrons in the $S_1$ should no longer be possible and we would expect the characteristic Gr phonons to become measurable in Raman spectroscopy.
We conducted \emph{ex situ} Raman spectroscopy for a sample with $(\sqrt{3} \times \sqrt{3})$R$30^{\circ}$ Eu intercalation structure and found indeed the characteristic Raman active Gr phonon modes to be present. However, significant complications arise in the interpretation of the Raman spectrum because of the high doping level, cf.\ Supplemental Material \cite{Supplement}.

%

In conclusion, the effect of a Eu intercalation layer on the Gr band structure is to first approximation a rigid band shift resulting in n-doping, a lifting of hybridization of Gr with the Ir $S_1$ surface state close to the Fermi level, and a strong reduction of the moiré superperiodic potential. 

\section{Summary}

We identified two well-ordered Eu intercalation layers, either with a $(2 \times 2)$ or a $(\sqrt{3} \times \sqrt{3})\mathrm{R}30^{\circ}$ superstructure with respect to Gr. Using PEEM and LEEM, we found intercalation to take place by penetration at wrinkles, if present.

The two intercalation structures exhibit fundamentally different magnetic behavior:
We have found the $(2 \times 2)$ layer to be simply paramagnetic. In contrast, the $(\sqrt{3} \times \sqrt{3})$R$30^{\circ}$ layer is ferromagnetically coupled, yet does not exhibit either spontaneous magnetization or hysteresis.
Easy-plane anisotropy is for both structures understood as shape anisotropy of the Eu monolayer.
Interestingly, the magnetic behavior of the $(\sqrt{3} \times \sqrt{3})$R$30^{\circ}$ layer strongly deviates from its bulk counterpart EuC$_6$ indicating the importance of the Ir substrate and the restriction to a single layer.

Using ARPES we find the band structure of Gr to stay largely intact upon Eu intercalation. The shift of the Gr bands induced by doping was determined to be about $\unit[-1.36]{eV}$ for the $(2 \times 2)$ and $\unit[-1.43]{eV}$ for the $(\sqrt{3} \times \sqrt{3})$R$30^{\circ}$ structure.
Additionaly, we find a shift of the Ir surface state $S_1$ to $\unit[-0.4]{eV}$ in the $(\sqrt{3} \times \sqrt{3})$R$30^{\circ}$ structure. Accordingly, hybridization between Gr states and Ir surface states is strongly suppressed. 

\section{Acknowledgements}
We thank Thomas Lorenz, Achim Rosch, and Heiko Wende for helpful discussions.
The authors acknowledge financial support through the Institutional Strategy of the University of Cologne
within the German Excellence Initiative and through the DFG Priority Program 1459 ``Graphene'' within project MI581/20-1. 


\clearpage
\begin{widetext}
\begin{center}
\textbf{\large Supplemental Material}
\end{center}
\setcounter{equation}{0}
\setcounter{figure}{0}
\setcounter{table}{0}
\makeatletter
\renewcommand{\theequation}{S\arabic{equation}}
\renewcommand{\thefigure}{S\arabic{figure}}

\section{Monitoring intercalation during deposition}

Figure~\ref{supplfig1}(a) shows an exemplary $\text{\textmu}$-LEED image of Gr on Ir(111) prior to Eu deposition, confirming the high quality of the initial layer. As apparent in the PEEM image in Fig.~\ref{supplfig1}(b), taken in the early stage of Eu deposition, the intercalation starts at a few distinct points separated by several \unit[100]{nm}. With increasing coverage, the intercalated Eu spreads in large islands [Figs.~\ref{supplfig1}(c)\,-\,(f)]. The preferred orientation from the top left to the bottom right is given by step bunches of the Ir(111) crystal, which act as diffusion barriers for the intercalated material. In the end, the Gr layer is fully intercalated by Eu [Fig.~\ref{supplfig1}(g)]. All over the sample, $\text{\textmu}$-LEED gives a $(\sqrt{3} \times \sqrt{3})$R$30^{\circ}$ diffraction pattern as shown in Fig.~\ref{supplfig1}(h).

\begin{figure} [htbp]
\begin{center}
\includegraphics[width=1.0\textwidth]{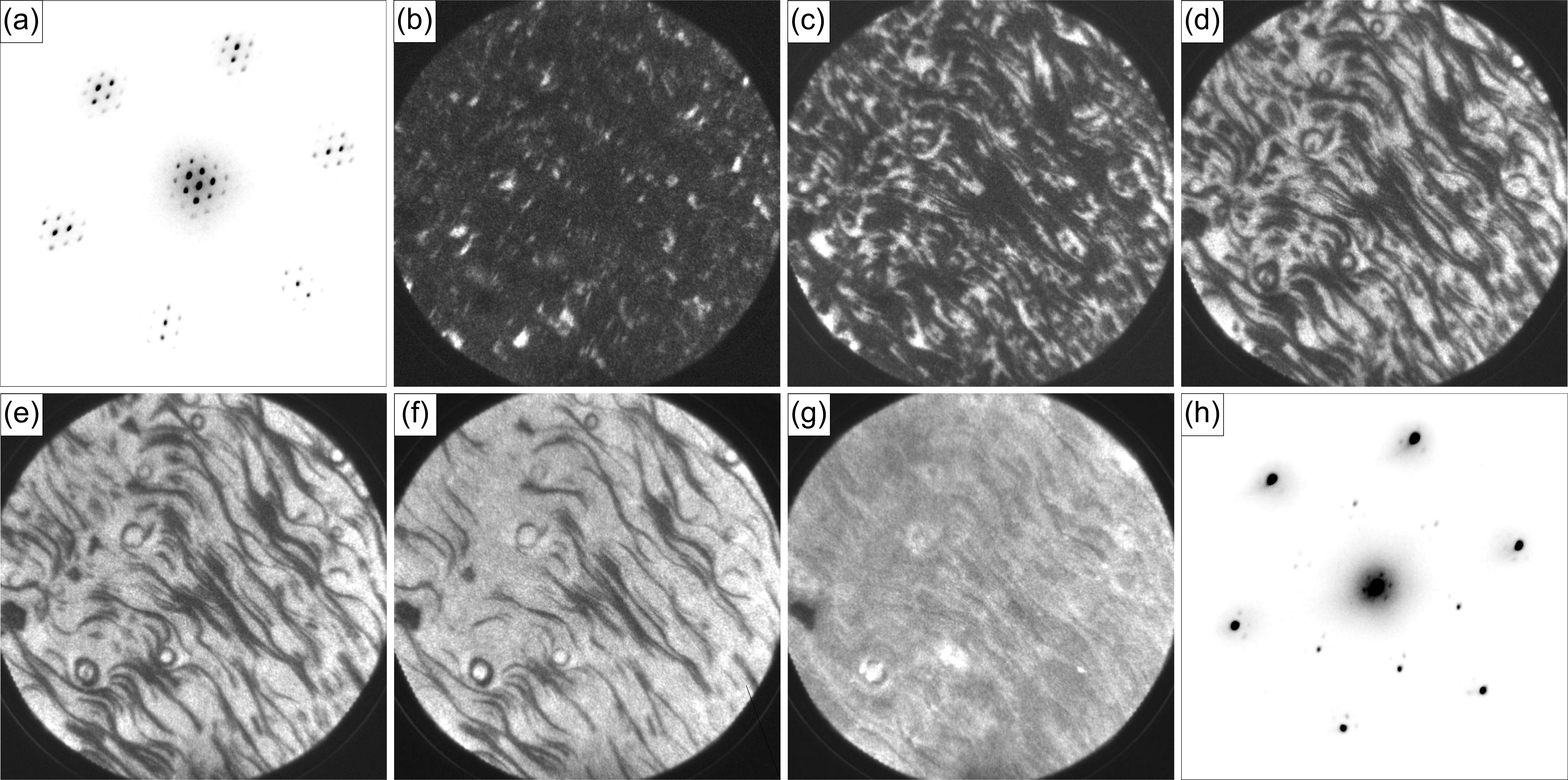}	
\caption{\label{supplfig1} 
(a) $\text{\textmu}$-LEED of pristine Gr (\unit[45]{eV} electron energy, illuminated area $\unit[1]{\text{\textmu}m}$ diameter). (b)\,-\,(g) PEEM series during Eu deposition at \unit[620]{K} starting with the nucleation of intercalation and ending with a fully saturated layer ($\unit[25]{\text{\textmu}m}$ field of view). (h) $\text{\textmu}$-LEED after complete intercalation showing the  $(\sqrt{3} \times \sqrt{3})$R$30^{\circ}$ superstructure (\unit[45]{eV} electron energy, illuminated area $\unit[1]{\text{\textmu}m}$ diameter).}
\end{center}
\end{figure}

\section{XAS and XMCD spectra of the \texorpdfstring{$(2 \times 2)$}{2x2} structure}

Figure~\ref{supplfig2} shows the XAS and XMCD spectra for the $(2 \times 2)$ intercalation structure in analogy to the data for the $(\sqrt{3} \times \sqrt{3})$R$30^{\circ}$ presented in the manuscript. The spectra are qualitatively similar, except for the smaller XMCD effect in the case of the $(2 \times 2)$ structure, indicating the smaller spin moment at \unit[5]{T}.

\begin{figure} [htbp]
\begin{center}
\includegraphics[width=0.5\textwidth]{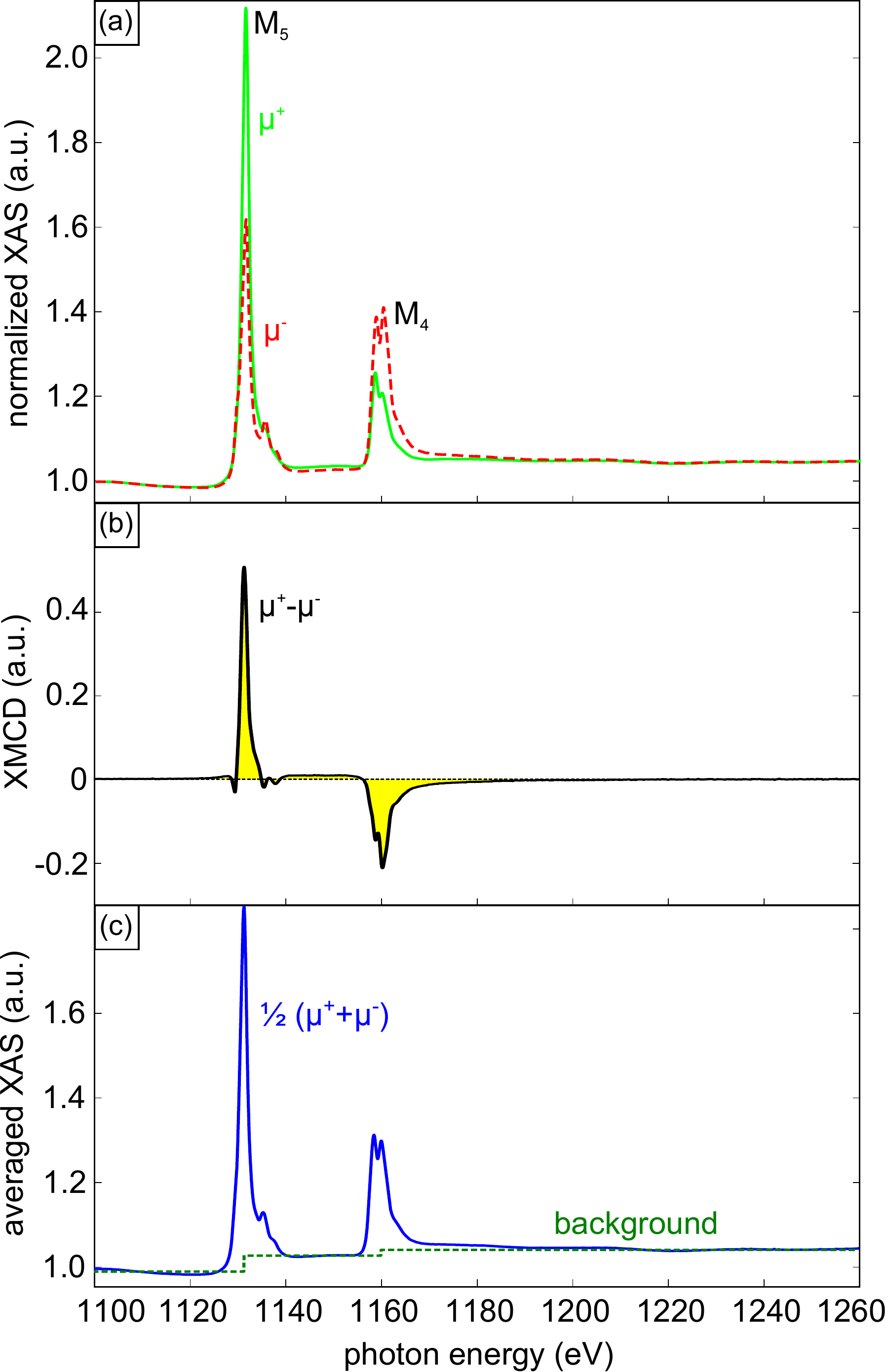}	
\caption{(a) Normalized XAS signal at \unit[10]{K} and $\unit[5]{T}$ for positive ($\mu^+$, solid green line) and negative ($\mu^-$, dashed red line) helicities in normal incidence. (b) Resulting XMCD signal ($\mu^+ - \mu^-$). (c) Polarization-averaged XAS spectrum  $\frac{1}{2}(\mu^+ + \mu^-$) (blue solid line) with step-like continuum background (green dashed line).}
\label{supplfig2} 
\end{center}
\end{figure}

\section{Calculation of shape anisotropy with discrete dipoles}
%

Shape anisotropy in general is a result of the magnetostatic interaction of the dipoles that are formed by the magnetic moments. For certain geometries (e.g., an infinitely extended plane), the demagnetization factor allows to conveniently describe the shape anisotropy. However, the description via a demagnetization factor 
relies on a continuum treatment.
As the magnetic moment is not actually homogeneously distributed, but localized at the magnetic atoms,
a treatment of the dipolar interaction in a continuum model is expected to give rise to increasing errors
when a ferromagnetic layer becomes thinner.
This is already apparent from the fact that for the calculation via the demagnetization factor it is necessary to know the \emph{volume} susceptibility and thus a layer thickness has to be assumed, which is an ill-defined quantity for a monatomic layer.
An approach via discrete dipoles is then to be preferred \cite{Johnson1996}.
Therefore, 
we have calculated a discrete-dipole model below for the $(\sqrt{3} \times \sqrt{3})$R$30^{\circ}$ layer. The calculations show that no significant error is introduced by the continuum treatment in our case.

To account for the almost ferromagnetic behavior, our model has the discrete dipoles aligned parallel to each other within a disc.
The geometry is depicted in Fig.~\ref{supplfig3}.

\begin{figure} [htbp]
\begin{center}
\includegraphics[width=0.25\columnwidth]{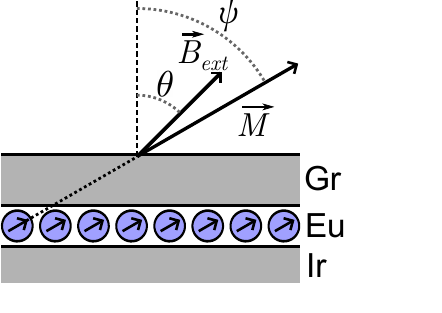}	
\caption{\label{supplfig3} 
(color online) Sketch of the situation assumed in the calculation of the shape anisotropy of the $(\sqrt{3} \times \sqrt{3})$R$30^{\circ}$ intercalation structure. The external magnetic field $\vec{B_{ext}}$ and the magnetization $\vec{M}$ form angles $\theta$ and $\psi$ with the surface normal, respectively.}
\end{center}
\end{figure}

According to basic magnetostatics the dipole-dipole interaction Hamiltonian of two magnetic moments $\vec{m}_i$, $\vec{m}_j$ separated by a distance vector $\vec{r}_{ij}$ is given by

\begin{equation}
H = - \frac{\mu_0} {4 \pi {|\vec{r}_{ij}|}^3 } \left( 3 (\vec{m}_i \cdot \vec{e}_{ij})  (\vec{m}_j \cdot \vec{e}_{ij}) - \vec{m}_i \cdot \vec{m}_j \right),
\end{equation}

where $\vec{e}_{ij}$ denotes the unit vector $\vec{e}_{ij}=\vec{r}_{ij}/|\vec{r}_{ij}|$. In the following we will assume for simplification ferromagnetic order of the spins, i.e., all spins are oriented in parallel. Hence, we have $\vec{m}_i=\vec{m}_j=\unit[7]{\mu_B} \cdot \vec{e}_{\vec{M}}$ with $\vec{e}_{\vec{M}}$ as the unit vector in the direction of magnetization $\vec{M}$. With that we can calculate the dipole-dipole interaction energy of one Eu atom with all others as a function of the magnetization direction:

\begin{equation}\label{eqn2}
E_{\mathrm{dipole}} = - \frac{1}{2}\sum_{\vec{r}_{ij} \neq \vec{0}}  \frac{ \mu_0 } {4 \pi {|\vec{r}_{ij}|}^3 } (\unit[7]{\mu_{\mathrm{B}}})^2 \left( 3 (\vec{e}_{\vec{M}} \cdot \vec{e}_{ij})^2 - 1 \right).
\end{equation}

The sum runs over the position vectors of Eu atoms

\begin{equation}
\vec{r}_{ij}= i \begin{pmatrix} 1 \\ 0 \\ 0 \end{pmatrix} + j \begin{pmatrix} -\frac{1}{2} \\ \sqrt{3} \\ 0 \end{pmatrix}  \cdot \sqrt{3} \cdot \unit[2.46]{\text{\AA}}.
\end{equation}

A factor of one half has been added to avoid double counting of pairwise interactions such that $E_{\mathrm{dipole}}$ can be properly understood as an anisotropy energy per Eu atom. $E_{\mathrm{dipole}}$ has been calculated for several magnetization directions by making a numerical summation over about $4\cdot 10^6$ neighbors. From this one can conclude on the following dependence of $E_{\mathrm{dipole}}$ on the angle $\psi$ between magnetization and surface normal:

\begin{equation}
E_{\mathrm{dipole}}=E_{\mathrm{aniso}} \cdot \left(-\frac{1}{3}+\cos^2\psi\right)
\label{eq:aniso}
\end{equation}

with $E_{\mathrm{aniso}} = \unit[281.3]{\text{\textmu} eV}$ per Eu atom.
We note that a continuum treatment, assuming a layer thickness equal to the Eu-Eu distance, leads to the estimate $E_{\mathrm{aniso}} = \unit[246]{\text{\textmu} eV}$ per Eu atom, which agrees reasonably well with the result of our discrete model.
The total energy in an external magnetic field of modulus $B$ oriented along an angle $\theta$ with respect to the surface normal is given by

\begin{equation}
E_\mathrm{total}=E_{\mathrm{aniso}}(-\frac{1}{3}+\cos^2{\psi}) - mB_\mathrm{ext}\cos(\psi-\theta) 
\end{equation}
with $m=\unit[7]{\mu_B}$.
The intuition of this equation is as follows: To minimize the anisotropy energy that results from dipolar interactions, the magnetization would prefer to be in-plane ($\psi=90^{\circ}$); this is captured by the anisotropy energy (first summand). However, the external field is able to pull the magnetization out of the plane by means of the additional Zeeman energy (second summand). So if we minimize $E_{\mathrm{total}}$ by setting its derivative to zero, we can calculate $\psi$ as a function of magnitude $B_\mathrm{ext}$ and angle $\theta$ of the external field:

\begin{equation}
\frac{\partial E_\mathrm{total}}{\partial \psi} = -2E_{\mathrm{aniso}}\sin{\psi}\cos{\psi} + mB_\mathrm{ext}\sin{(\psi-\theta)}\overset{!}{=}0
\end{equation}

Within this model, the zero-field susceptibility is infinite for a non-vanishing in-plane component of the magnetic field ($\theta \neq 0$) as there is no barrier for rotating the spins within the plane. In contrast, for $\theta=0$ the zero-field susceptibility is finite and can be matched to our normal incidence measurement. We obtain the simple relation

\begin{equation}
\cos{\psi}=\frac{mB_\mathrm{ext}}{2E_{\mathrm{aniso}}}.
\end{equation}

The observable in our experiment is the projection of the magnetic moment on the beam, which is $m \cos{\psi}$ for $\theta=0$. From this we can calculate the theoretical expectation for the zero-field susceptibility as

\begin{equation}
\chi_{\mathrm{theo}} = \frac{\partial}{\partial B_\mathrm{ext}} m \cos{\psi} = \frac{m^2}{2E_{\mathrm{aniso}}} =\unit[5.04]{\mu_B/T}.
\end{equation}

This value is quite close to the experimentally determined zero-field susceptibility for normal incidence of $\chi_\mathrm{exp}=\unit[6.2]{\mu_B/T}$, and deviates only around 15\% from the value of $\chi_\mathrm{exp}=\unit[5.8]{\mu_B/T}$ obtained via a continuum treatment. 

Lastly, we note that the the summation of Eq.~\ref{eqn2} converges quickly, reaching a value within a factor of two of the final value already for summation over only two nearest neighbors, i.e., 19 atoms arranged in a small hexagon. This indicates that the shape anisotropy is not significantly lower, if the ferromagnetic order is not long-range.
\section{Geometry used for the ARPES measurements}

Figure~\ref{supplfig4}(a) shows the first Brillouin zone of Gr with the high symmetry-points $\Gamma$, K, K' and M. The red line indicates the path of an ARPES spectrum along $\Gamma$KM direction with a small azimuthal misalignment $\Delta \phi$. Figure~\ref{supplfig4}(b) shows that the band structure is consequently not exactly cut across the K point. Thus, the spectra seemingly show a band gap.  

\begin{figure} [htbp]
\begin{center}
\includegraphics[width=0.6\textwidth]{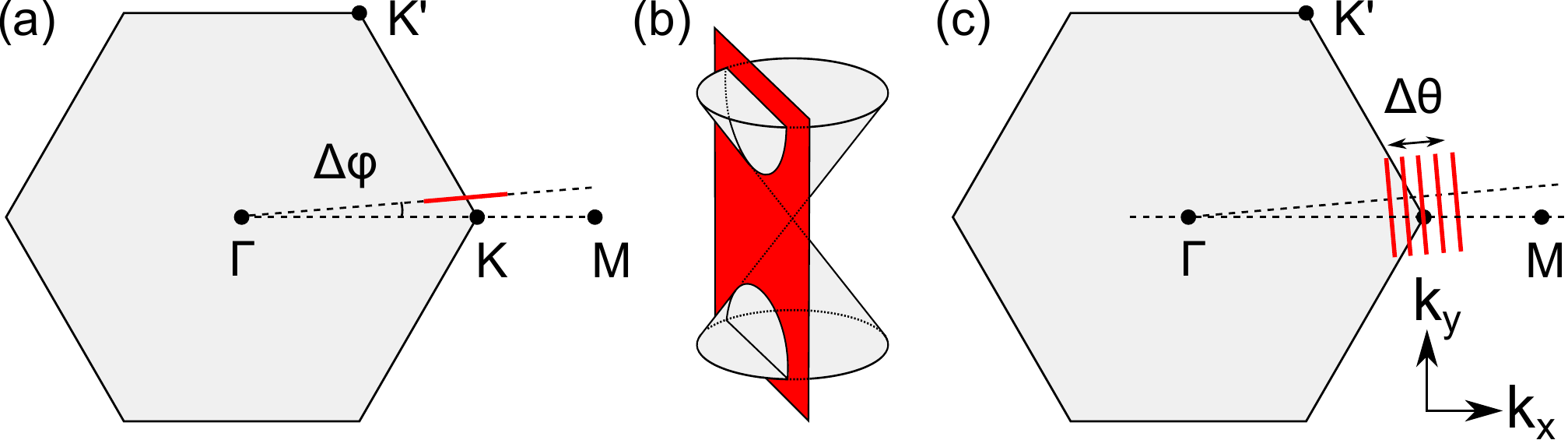}	
\caption{(a) ARPES measurement in $\Gamma$KM direction with an azimuthal offset $\Delta \phi$. (b) $\Delta \phi \neq 0$ leads to an apparent band gap at the K point. (c) Measurement perpendicular to the $\Gamma$KM direction for $\Delta \phi \neq 0$. By finely adjusting the polar angle $\theta$ the K point can be reached.}
\label{supplfig4} 
\end{center}
\end{figure}

When rotating the electron analyzer by $90^{\circ}$, maps perpendicular to the $\Gamma$KM direction can be taken. By changing the finely adjustable polar angle $\theta$ in small steps, the K point can be reached even in the presence of a small $\Delta \phi$. The corresponding geometry is shown in Fig.~\ref{supplfig4}(c). 

\section{\emph{Ex situ} Raman spectroscopy of Eu-intercalated Gr on Ir(111)}

Raman scattering of graphene phonon modes is a versatile tool to study the properties of graphene because the details of
the Raman spectrum depend on, e.g., the number of graphene layers, the doping level, strain, and disorder
(for a recent review, see Ref.~\onlinecite{Ferrari13}).
In the case of Gr/Ir(111), the typical graphene Raman peaks are absent if the lattice vectors of graphene and Ir(111) are aligned but the Raman features become visible if the graphene is rotated by 30$^\circ$~\cite{Starodub2011}.
Obviously, the Raman features are sensitive to the interaction between graphene and substrate.
It therefore seems worthwhile to analyze the Raman spectrum of Gr/Eu/Ir(111).

After intercalation of a ($\sqrt{3} \times \sqrt{3}$)R30$^\circ$ Eu layer under graphene on Ir(111), the sample was transferred
from the vacuum system to ambient conditions and {\it ex situ} Raman spectroscopy was conducted in an Ar-purged setup.
The data were measured for a scattering angle of about 90$^\circ$ using a laser with wavelength of 532~nm (2.33~eV).
The total incident power of 100~mW was focused on a spot with a diameter of about 100~$\mu$m.
The elastically scattered contribution was suppressed by an edge filter, whereas the inelastically scattered light was detected
using a grating with 600~grooves/mm and a cooled CCD.\@
The data shown in Fig.\ 5 have been averaged over 10 individual runs of 180~s each and smoothed by 3-point averaging.
Finally, a background has been subtracted.

The relatively strong intensity of Raman scattering from a single layer of, e.g., micromechanically cleaved graphene
is based on two effects.
First, the choice of an appropriate substrate such as 300~nm SiO$_2$ on Si gives rise to so-called
interference-enhanced Raman scattering (IERS). This does not apply to Gr/Eu/Ir(111), thus we expect
a smaller Raman signal.
Second, the Raman intensity is strongly enhanced in {\it resonant} Raman scattering,  in which case the incident
photon energy is in resonance with an excitation energy of the system. Due to the gapless structure of the Dirac cone,
the resonance condition is always fulfilled, at least for low-doped graphene. According to the ARPES data, the Dirac point
in the ($\sqrt{3} \times \sqrt{3}$)R30$^\circ$ sample is 1.43~eV below the Fermi energy. Vertical transitions between
the $\pi$ and $\pi^*$ bands thus require an energy of at least 2.86~eV.\@ In other words, the resonance condition
cannot be fulfilled for a laser energy of 2.33~eV.\@ Accordingly, we do not expect to see a Raman signal from
Gr/Eu/Ir(111) due to the high doping level.

In contrast to this expectation, Fig.~\ref{supplfig5} shows the typical Raman features of defective graphene \cite{Ferrari13}.
We attribute the peaks at 1597~cm$^{-1}$ and 2679~cm$^{-1}$ to the Raman-active G and G' peaks, respectively,
while the peaks at 1352~cm$^{-1}$ (D peak) and at about 2450~cm$^{-1}$ (D+D'') can be interpreted as defect-induced
modes (cf.\ Ref.\ [\onlinecite{Ferrari13}]).
Due to the high reactivity of Eu, we cannot exclude that defects are created under ambient conditions.
The observation of these modes for a laser energy of 2.33~eV can be explained under the assumption
that the doping level is lower under ambient conditions, thus the resonance condition can be fulfilled.
Such a reduction of doping could result from adsorption to the frontside of graphene.
The peak positions are consistent with the assumption of a reduced doping level.
The G (G') peak is observed at 1580~cm$^{-1}$ (2690~cm$^{-1}$) in pristine graphene \cite{Ferrari13}.
Doping levels equivalent to shifts of the Fermi energy of almost 1~eV were obtained
by electrochemical doping \cite{Das08,Kalbac10,Chen11}.
For such high levels of electron doping, the G (G') peak shifts to higher (lower) energy.
The peak energies depicted in Fig.\ 5 roughly agree with the data of Das {\it et al.} \cite{Das08}
for a Fermi energy located at about 0.6 - 0.7~eV above the Dirac point.\@
However, Raman peak energies also depend on the laser energy and on strain,
invalidating a direct quantitative comparison with the results of Das and collaborators.

Resonant Raman scattering is suppressed by Pauli blocking if the energy of the incident photons is
smaller than half the Fermi energy. In case of the G peak, it has been show that Pauli blocking counterintuitively
may yield an {\it enhanced} Raman intensity because it removes the destructive interference of different Raman
channels~\cite{Chen11,Basko09}.
This does not apply to the G' peak, which shows a strong decrease of the intensity upon Pauli blocking~\cite{Chen11,Zhao11}.
Therefore, the Raman signal shown in Fig.~\ref{supplfig5} has to originate from sample areas where the Fermi energy is smaller
than about $1.1$~eV.


\begin{figure} [htbp]
\begin{center}
\includegraphics[width=0.6\textwidth]{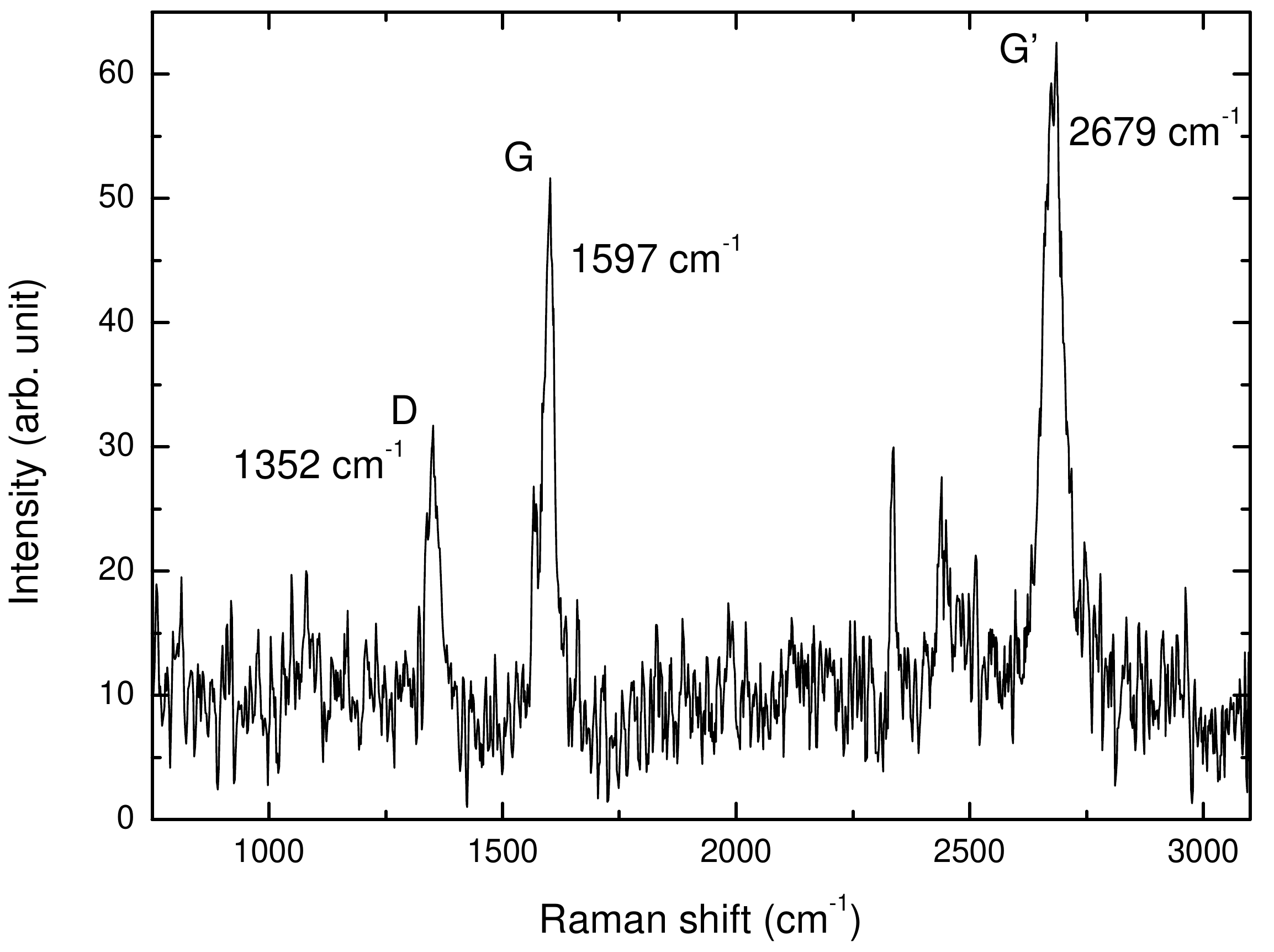}	
\caption{\emph{Ex situ} Raman spectrum of Gr on Ir(111) intercalated by a $(\sqrt{3} \times \sqrt{3})$R$30^{\circ}$ Eu layer. The positions of the Raman-active Gr phonon peaks G and G' as well as of the disorder-induced peak D are indicated.}
\label{supplfig5} 
\end{center}
\end{figure}

\end{widetext}
\clearpage

\bibliography{references}

\begin{thebibliography}{67}%
\makeatletter
\providecommand \@ifxundefined [1]{%
 \@ifx{#1\undefined}
}%
\providecommand \@ifnum [1]{%
 \ifnum #1\expandafter \@firstoftwo
 \else \expandafter \@secondoftwo
 \fi
}%
\providecommand \@ifx [1]{%
 \ifx #1\expandafter \@firstoftwo
 \else \expandafter \@secondoftwo
 \fi
}%
\providecommand \natexlab [1]{#1}%
\providecommand \enquote  [1]{``#1''}%
\providecommand \bibnamefont  [1]{#1}%
\providecommand \bibfnamefont [1]{#1}%
\providecommand \citenamefont [1]{#1}%
\providecommand \href@noop [0]{\@secondoftwo}%
\providecommand \href [0]{\begingroup \@sanitize@url \@href}%
\providecommand \@href[1]{\@@startlink{#1}\@@href}%
\providecommand \@@href[1]{\endgroup#1\@@endlink}%
\providecommand \@sanitize@url [0]{\catcode `\\12\catcode `\$12\catcode
  `\&12\catcode `\#12\catcode `\^12\catcode `\_12\catcode `\%12\relax}%
\providecommand \@@startlink[1]{}%
\providecommand \@@endlink[0]{}%
\providecommand \url  [0]{\begingroup\@sanitize@url \@url }%
\providecommand \@url [1]{\endgroup\@href {#1}{\urlprefix }}%
\providecommand \urlprefix  [0]{URL }%
\providecommand \Eprint [0]{\href }%
\providecommand \doibase [0]{http://dx.doi.org/}%
\providecommand \selectlanguage [0]{\@gobble}%
\providecommand \bibinfo  [0]{\@secondoftwo}%
\providecommand \bibfield  [0]{\@secondoftwo}%
\providecommand \translation [1]{[#1]}%
\providecommand \BibitemOpen [0]{}%
\providecommand \bibitemStop [0]{}%
\providecommand \bibitemNoStop [0]{.\EOS\space}%
\providecommand \EOS [0]{\spacefactor3000\relax}%
\providecommand \BibitemShut  [1]{\csname bibitem#1\endcsname}%
\let\auto@bib@innerbib\@empty
\bibitem [{\citenamefont {Min}\ \emph {et~al.}(2006)\citenamefont {Min},
  \citenamefont {Hill}, \citenamefont {Sinitsyn}, \citenamefont {Sahu},
  \citenamefont {Kleinman},\ and\ \citenamefont {MacDonald}}]{Min2006}%
  \BibitemOpen
  \bibfield  {author} {\bibinfo {author} {\bibfnamefont {H.}~\bibnamefont
  {Min}}, \bibinfo {author} {\bibfnamefont {J.~E.}\ \bibnamefont {Hill}},
  \bibinfo {author} {\bibfnamefont {N.~A.}\ \bibnamefont {Sinitsyn}}, \bibinfo
  {author} {\bibfnamefont {B.~R.}\ \bibnamefont {Sahu}}, \bibinfo {author}
  {\bibfnamefont {L.}~\bibnamefont {Kleinman}}, \ and\ \bibinfo {author}
  {\bibfnamefont {A.~H.}\ \bibnamefont {MacDonald}},\ }\href {\doibase
  10.1103/PhysRevB.74.165310} {\bibfield  {journal} {\bibinfo  {journal} {Phys.
  Rev. B}\ }\textbf {\bibinfo {volume} {74}},\ \bibinfo {pages} {165310}
  (\bibinfo {year} {2006})}\BibitemShut {NoStop}%
\bibitem [{\citenamefont {Yazyev}(2008)}]{Yazyev2008}%
  \BibitemOpen
  \bibfield  {author} {\bibinfo {author} {\bibfnamefont {O.~V.}\ \bibnamefont
  {Yazyev}},\ }\href {\doibase 10.1021/nl072667q} {\bibfield  {journal}
  {\bibinfo  {journal} {Nano Lett.}\ }\textbf {\bibinfo {volume} {8}},\
  \bibinfo {pages} {1011} (\bibinfo {year} {2008})}\BibitemShut {NoStop}%
\bibitem [{\citenamefont {Varykhalov}\ and\ \citenamefont
  {Rader}(2009)}]{Varykhalov2009}%
  \BibitemOpen
  \bibfield  {author} {\bibinfo {author} {\bibfnamefont {A.}~\bibnamefont
  {Varykhalov}}\ and\ \bibinfo {author} {\bibfnamefont {O.}~\bibnamefont
  {Rader}},\ }\href {\doibase 10.1103/PhysRevB.80.035437} {\bibfield  {journal}
  {\bibinfo  {journal} {Phys. Rev. B}\ }\textbf {\bibinfo {volume} {80}},\
  \bibinfo {pages} {035437} (\bibinfo {year} {2009})}\BibitemShut {NoStop}%
\bibitem [{\citenamefont {Weser}\ \emph {et~al.}(2011)\citenamefont {Weser},
  \citenamefont {Voloshina}, \citenamefont {Horn},\ and\ \citenamefont
  {Dedkov}}]{Weser2011}%
  \BibitemOpen
  \bibfield  {author} {\bibinfo {author} {\bibfnamefont {M.}~\bibnamefont
  {Weser}}, \bibinfo {author} {\bibfnamefont {E.~N.}\ \bibnamefont
  {Voloshina}}, \bibinfo {author} {\bibfnamefont {K.}~\bibnamefont {Horn}}, \
  and\ \bibinfo {author} {\bibfnamefont {Y.~S.}\ \bibnamefont {Dedkov}},\
  }\href {\doibase 10.1039/C1CP00014D} {\bibfield  {journal} {\bibinfo
  {journal} {Phys. Chem. Chem. Phys.}\ }\textbf {\bibinfo {volume} {13}},\
  \bibinfo {pages} {7534} (\bibinfo {year} {2011})}\BibitemShut {NoStop}%
\bibitem [{\citenamefont {Haugen}\ \emph {et~al.}(2008)\citenamefont {Haugen},
  \citenamefont {Huertas-Hernando},\ and\ \citenamefont
  {Brataas}}]{Haugen2008}%
  \BibitemOpen
  \bibfield  {author} {\bibinfo {author} {\bibfnamefont {H.}~\bibnamefont
  {Haugen}}, \bibinfo {author} {\bibfnamefont {D.}~\bibnamefont
  {Huertas-Hernando}}, \ and\ \bibinfo {author} {\bibfnamefont
  {A.}~\bibnamefont {Brataas}},\ }\href {\doibase 10.1103/PhysRevB.77.115406}
  {\bibfield  {journal} {\bibinfo  {journal} {Phys. Rev. B}\ }\textbf {\bibinfo
  {volume} {77}},\ \bibinfo {pages} {115406} (\bibinfo {year}
  {2008})}\BibitemShut {NoStop}%
\bibitem [{\citenamefont {Klinkhammer}\ \emph {et~al.}(2013)\citenamefont
  {Klinkhammer}, \citenamefont {F\"orster}, \citenamefont {Schumacher},
  \citenamefont {Oepen}, \citenamefont {Michely},\ and\ \citenamefont
  {Busse}}]{Klinkhammer2013}%
  \BibitemOpen
  \bibfield  {author} {\bibinfo {author} {\bibfnamefont {J.}~\bibnamefont
  {Klinkhammer}}, \bibinfo {author} {\bibfnamefont {D.~F.}\ \bibnamefont
  {F\"orster}}, \bibinfo {author} {\bibfnamefont {S.}~\bibnamefont
  {Schumacher}}, \bibinfo {author} {\bibfnamefont {H.~P.}\ \bibnamefont
  {Oepen}}, \bibinfo {author} {\bibfnamefont {T.}~\bibnamefont {Michely}}, \
  and\ \bibinfo {author} {\bibfnamefont {C.}~\bibnamefont {Busse}},\ }\href
  {\doibase 10.1063/1.4821953} {\bibfield  {journal} {\bibinfo  {journal}
  {Appl. Phys. Lett.}\ }\textbf {\bibinfo {volume} {103}},\ \bibinfo {pages}
  {131601} (\bibinfo {year} {2013})}\BibitemShut {NoStop}%
\bibitem [{\citenamefont {Klinkhammer}\ \emph {et~al.}(2014)\citenamefont
  {Klinkhammer}, \citenamefont {Schlipf}, \citenamefont {Craes}, \citenamefont
  {Runte}, \citenamefont {Michely},\ and\ \citenamefont
  {Busse}}]{Klinkhammer2014}%
  \BibitemOpen
  \bibfield  {author} {\bibinfo {author} {\bibfnamefont {J.}~\bibnamefont
  {Klinkhammer}}, \bibinfo {author} {\bibfnamefont {M.}~\bibnamefont
  {Schlipf}}, \bibinfo {author} {\bibfnamefont {F.}~\bibnamefont {Craes}},
  \bibinfo {author} {\bibfnamefont {S.}~\bibnamefont {Runte}}, \bibinfo
  {author} {\bibfnamefont {T.}~\bibnamefont {Michely}}, \ and\ \bibinfo
  {author} {\bibfnamefont {C.}~\bibnamefont {Busse}},\ }\href {\doibase
  10.1103/PhysRevLett.112.016803} {\bibfield  {journal} {\bibinfo  {journal}
  {Phys. Rev. Lett.}\ }\textbf {\bibinfo {volume} {112}},\ \bibinfo {pages}
  {016803} (\bibinfo {year} {2014})}\BibitemShut {NoStop}%
\bibitem [{\citenamefont {Olsen}\ \emph {et~al.}(1962)\citenamefont {Olsen},
  \citenamefont {Nereson},\ and\ \citenamefont {Arnold}}]{Olsen1962}%
  \BibitemOpen
  \bibfield  {author} {\bibinfo {author} {\bibfnamefont {C.~E.}\ \bibnamefont
  {Olsen}}, \bibinfo {author} {\bibfnamefont {N.~G.}\ \bibnamefont {Nereson}},
  \ and\ \bibinfo {author} {\bibfnamefont {G.~P.}\ \bibnamefont {Arnold}},\
  }\href {\doibase 10.1063/1.1728632} {\bibfield  {journal} {\bibinfo
  {journal} {J. Appl. Phys.}\ }\textbf {\bibinfo {volume} {33}},\ \bibinfo
  {pages} {1135} (\bibinfo {year} {1962})}\BibitemShut {NoStop}%
\bibitem [{\citenamefont {Nereson}\ \emph {et~al.}(1964)\citenamefont
  {Nereson}, \citenamefont {Olsen},\ and\ \citenamefont
  {Arnold}}]{Nereson1964}%
  \BibitemOpen
  \bibfield  {author} {\bibinfo {author} {\bibfnamefont {N.~G.}\ \bibnamefont
  {Nereson}}, \bibinfo {author} {\bibfnamefont {C.~E.}\ \bibnamefont {Olsen}},
  \ and\ \bibinfo {author} {\bibfnamefont {G.~P.}\ \bibnamefont {Arnold}},\
  }\href {\doibase 10.1103/PhysRev.135.A176} {\bibfield  {journal} {\bibinfo
  {journal} {Phys. Rev.}\ }\textbf {\bibinfo {volume} {135}},\ \bibinfo {pages}
  {176} (\bibinfo {year} {1964})}\BibitemShut {NoStop}%
\bibitem [{\citenamefont {Arnold}\ \emph {et~al.}(1964)\citenamefont {Arnold},
  \citenamefont {Olsen},\ and\ \citenamefont {Nereson}}]{Arnold1964}%
  \BibitemOpen
  \bibfield  {author} {\bibinfo {author} {\bibfnamefont {G.~P.}\ \bibnamefont
  {Arnold}}, \bibinfo {author} {\bibfnamefont {C.~E.}\ \bibnamefont {Olsen}}, \
  and\ \bibinfo {author} {\bibfnamefont {N.~G.}\ \bibnamefont {Nereson}},\
  }\href {\doibase 10.1063/1.1713365} {\bibfield  {journal} {\bibinfo
  {journal} {J. Appl. Phys.}\ }\textbf {\bibinfo {volume} {35}},\ \bibinfo
  {pages} {1031} (\bibinfo {year} {1964})}\BibitemShut {NoStop}%
\bibitem [{\citenamefont {F\"{o}rster}\ \emph {et~al.}(2012)\citenamefont
  {F\"{o}rster}, \citenamefont {Wehling}, \citenamefont {Schumacher},
  \citenamefont {Rosch},\ and\ \citenamefont {Michely}}]{Forster2012b}%
  \BibitemOpen
  \bibfield  {author} {\bibinfo {author} {\bibfnamefont {D.~F.}\ \bibnamefont
  {F\"{o}rster}}, \bibinfo {author} {\bibfnamefont {T.~O.}\ \bibnamefont
  {Wehling}}, \bibinfo {author} {\bibfnamefont {S.}~\bibnamefont {Schumacher}},
  \bibinfo {author} {\bibfnamefont {A.}~\bibnamefont {Rosch}}, \ and\ \bibinfo
  {author} {\bibfnamefont {T.}~\bibnamefont {Michely}},\ }\href {\doibase
  10.1088/1367-2630/14/2/023022} {\bibfield  {journal} {\bibinfo  {journal}
  {New J. Phys.}\ }\textbf {\bibinfo {volume} {14}},\ \bibinfo {pages} {23022}
  (\bibinfo {year} {2012})}\BibitemShut {NoStop}%
\bibitem [{\citenamefont {Schumacher}\ \emph
  {et~al.}(2013{\natexlab{a}})\citenamefont {Schumacher}, \citenamefont
  {Förster}, \citenamefont {Rösner}, \citenamefont {Wehling},\ and\
  \citenamefont {Michely}}]{Schumacher2013a}%
  \BibitemOpen
  \bibfield  {author} {\bibinfo {author} {\bibfnamefont {S.}~\bibnamefont
  {Schumacher}}, \bibinfo {author} {\bibfnamefont {D.~F.}\ \bibnamefont
  {Förster}}, \bibinfo {author} {\bibfnamefont {M.}~\bibnamefont {Rösner}},
  \bibinfo {author} {\bibfnamefont {T.~O.}\ \bibnamefont {Wehling}}, \ and\
  \bibinfo {author} {\bibfnamefont {T.}~\bibnamefont {Michely}},\ }\href
  {\doibase 10.1103/PhysRevLett.110.086111} {\bibfield  {journal} {\bibinfo
  {journal} {Phys. Rev. Lett.}\ }\textbf {\bibinfo {volume} {110}},\ \bibinfo
  {pages} {086111} (\bibinfo {year} {2013}{\natexlab{a}})}\BibitemShut
  {NoStop}%
\bibitem [{\citenamefont {Schumacher}\ \emph
  {et~al.}(2013{\natexlab{b}})\citenamefont {Schumacher}, \citenamefont
  {Wehling}, \citenamefont {Lazi{\'c}}, \citenamefont {Runte}, \citenamefont
  {F{\"o}rster}, \citenamefont {Busse}, \citenamefont {Petrovi{\'c}},
  \citenamefont {Kralj}, \citenamefont {Bl{\"u}gel}, \citenamefont
  {Atodiresei}, \citenamefont {Caciuc},\ and\ \citenamefont
  {Michely}}]{Schumacher2013b}%
  \BibitemOpen
  \bibfield  {author} {\bibinfo {author} {\bibfnamefont {S.}~\bibnamefont
  {Schumacher}}, \bibinfo {author} {\bibfnamefont {T.~O.}\ \bibnamefont
  {Wehling}}, \bibinfo {author} {\bibfnamefont {P.}~\bibnamefont {Lazi{\'c}}},
  \bibinfo {author} {\bibfnamefont {S.}~\bibnamefont {Runte}}, \bibinfo
  {author} {\bibfnamefont {D.~F.}\ \bibnamefont {F{\"o}rster}}, \bibinfo
  {author} {\bibfnamefont {C.}~\bibnamefont {Busse}}, \bibinfo {author}
  {\bibfnamefont {M.}~\bibnamefont {Petrovi{\'c}}}, \bibinfo {author}
  {\bibfnamefont {M.}~\bibnamefont {Kralj}}, \bibinfo {author} {\bibfnamefont
  {S.}~\bibnamefont {Bl{\"u}gel}}, \bibinfo {author} {\bibfnamefont
  {N.}~\bibnamefont {Atodiresei}}, \bibinfo {author} {\bibfnamefont
  {V.}~\bibnamefont {Caciuc}}, \ and\ \bibinfo {author} {\bibfnamefont
  {T.}~\bibnamefont {Michely}},\ }\href {\doibase 10.1021/nl402797j} {\bibfield
   {journal} {\bibinfo  {journal} {Nano Lett.}\ }\textbf {\bibinfo {volume}
  {13}},\ \bibinfo {pages} {5013} (\bibinfo {year}
  {2013}{\natexlab{b}})}\BibitemShut {NoStop}%
\bibitem [{\citenamefont {Suematsu}\ \emph {et~al.}(1983)\citenamefont
  {Suematsu}, \citenamefont {Ohmatsu}, \citenamefont {Sakakibara},
  \citenamefont {Date},\ and\ \citenamefont {Suzuki}}]{Suematsu1983}%
  \BibitemOpen
  \bibfield  {author} {\bibinfo {author} {\bibfnamefont {H.}~\bibnamefont
  {Suematsu}}, \bibinfo {author} {\bibfnamefont {K.}~\bibnamefont {Ohmatsu}},
  \bibinfo {author} {\bibfnamefont {T.}~\bibnamefont {Sakakibara}}, \bibinfo
  {author} {\bibfnamefont {M.}~\bibnamefont {Date}}, \ and\ \bibinfo {author}
  {\bibfnamefont {M.}~\bibnamefont {Suzuki}},\ }\href {\doibase
  10.1016/0038-1098(81)90966-2} {\bibfield  {journal} {\bibinfo  {journal}
  {Synthetic Met.}\ }\textbf {\bibinfo {volume} {8}},\ \bibinfo {pages} {23}
  (\bibinfo {year} {1983})}\BibitemShut {NoStop}%
\bibitem [{\citenamefont {Chen}\ \emph {et~al.}(1986)\citenamefont {Chen},
  \citenamefont {Dresselhaus}, \citenamefont {Dresselhaus}, \citenamefont
  {Suematsu}, \citenamefont {Minemoto}, \citenamefont {Ohmatsu},\ and\
  \citenamefont {Yosida}}]{Chen1986}%
  \BibitemOpen
  \bibfield  {author} {\bibinfo {author} {\bibfnamefont {S.~T.}\ \bibnamefont
  {Chen}}, \bibinfo {author} {\bibfnamefont {M.~S.}\ \bibnamefont
  {Dresselhaus}}, \bibinfo {author} {\bibfnamefont {G.}~\bibnamefont
  {Dresselhaus}}, \bibinfo {author} {\bibfnamefont {H.}~\bibnamefont
  {Suematsu}}, \bibinfo {author} {\bibfnamefont {H.}~\bibnamefont {Minemoto}},
  \bibinfo {author} {\bibfnamefont {K.}~\bibnamefont {Ohmatsu}}, \ and\
  \bibinfo {author} {\bibfnamefont {Y.}~\bibnamefont {Yosida}},\ }\href
  {\doibase 10.1103/PhysRevB.34.423} {\bibfield  {journal} {\bibinfo  {journal}
  {Phys. Rev. B}\ }\textbf {\bibinfo {volume} {34}},\ \bibinfo {pages} {423}
  (\bibinfo {year} {1986})}\BibitemShut {NoStop}%
\bibitem [{\citenamefont {{van Gastel}}\ \emph {et~al.}(2009)\citenamefont
  {{van Gastel}}, \citenamefont {N'Diaye}, \citenamefont {Wall}, \citenamefont
  {Coraux}, \citenamefont {Busse}, \citenamefont {Buckanie}, \citenamefont
  {{Meyer zu Heringdorf}}, \citenamefont {{Horn-von Hoegen}}, \citenamefont
  {Michely},\ and\ \citenamefont {Poelsema}}]{vanGastel2009}%
  \BibitemOpen
  \bibfield  {author} {\bibinfo {author} {\bibfnamefont {R.}~\bibnamefont {{van
  Gastel}}}, \bibinfo {author} {\bibfnamefont {A.~T.}\ \bibnamefont {N'Diaye}},
  \bibinfo {author} {\bibfnamefont {D.}~\bibnamefont {Wall}}, \bibinfo {author}
  {\bibfnamefont {J.}~\bibnamefont {Coraux}}, \bibinfo {author} {\bibfnamefont
  {C.}~\bibnamefont {Busse}}, \bibinfo {author} {\bibfnamefont {N.~M.}\
  \bibnamefont {Buckanie}}, \bibinfo {author} {\bibfnamefont {F.-J.}\
  \bibnamefont {{Meyer zu Heringdorf}}}, \bibinfo {author} {\bibfnamefont
  {M.}~\bibnamefont {{Horn-von Hoegen}}}, \bibinfo {author} {\bibfnamefont
  {T.}~\bibnamefont {Michely}}, \ and\ \bibinfo {author} {\bibfnamefont
  {B.}~\bibnamefont {Poelsema}},\ }\href {\doibase 10.1063/1.3225554}
  {\bibfield  {journal} {\bibinfo  {journal} {Appl. Phys. Lett.}\ }\textbf
  {\bibinfo {volume} {95}},\ \bibinfo {pages} {121901} (\bibinfo {year}
  {2009})}\BibitemShut {NoStop}%
\bibitem [{Ame()}]{Ames}%
  \BibitemOpen
  \href@noop {} {}\bibinfo {note} {Ames Laboratory (Materials Preparation
  Center) of the US DOE, Iowa State University, Ames, IA
  50011-3020}\BibitemShut {NoStop}%
\bibitem [{\citenamefont {Horcas}\ \emph {et~al.}(2007)\citenamefont {Horcas},
  \citenamefont {Fernández}, \citenamefont {Gómez-Rodríguez}, \citenamefont
  {Colchero}, \citenamefont {Gómez-Herrero},\ and\ \citenamefont
  {Baro}}]{Horcas2007}%
  \BibitemOpen
  \bibfield  {author} {\bibinfo {author} {\bibfnamefont {I.}~\bibnamefont
  {Horcas}}, \bibinfo {author} {\bibfnamefont {R.}~\bibnamefont {Fernández}},
  \bibinfo {author} {\bibfnamefont {J.~M.}\ \bibnamefont {Gómez-Rodríguez}},
  \bibinfo {author} {\bibfnamefont {J.}~\bibnamefont {Colchero}}, \bibinfo
  {author} {\bibfnamefont {J.}~\bibnamefont {Gómez-Herrero}}, \ and\ \bibinfo
  {author} {\bibfnamefont {A.~M.}\ \bibnamefont {Baro}},\ }\href {\doibase
  10.1063/1.2432410} {\bibfield  {journal} {\bibinfo  {journal} {Rev. Sci.
  Instrum.}\ }\textbf {\bibinfo {volume} {78}},\ \bibinfo {pages} {013705}
  (\bibinfo {year} {2007})}\BibitemShut {NoStop}%
\bibitem [{\citenamefont {Schmidt}\ \emph {et~al.}(1998)\citenamefont
  {Schmidt}, \citenamefont {Heun}, \citenamefont {Slezak}, \citenamefont
  {Diaz}, \citenamefont {Prince}, \citenamefont {Lilienkamp},\ and\
  \citenamefont {Bauer}}]{Schmidt1998}%
  \BibitemOpen
  \bibfield  {author} {\bibinfo {author} {\bibfnamefont {T.}~\bibnamefont
  {Schmidt}}, \bibinfo {author} {\bibfnamefont {S.}~\bibnamefont {Heun}},
  \bibinfo {author} {\bibfnamefont {J.}~\bibnamefont {Slezak}}, \bibinfo
  {author} {\bibfnamefont {J.}~\bibnamefont {Diaz}}, \bibinfo {author}
  {\bibfnamefont {K.~C.}\ \bibnamefont {Prince}}, \bibinfo {author}
  {\bibfnamefont {G.}~\bibnamefont {Lilienkamp}}, \ and\ \bibinfo {author}
  {\bibfnamefont {E.}~\bibnamefont {Bauer}},\ }\href {\doibase
  10.1142/S0218625X98001626} {\bibfield  {journal} {\bibinfo  {journal} {Surf.
  Rev. Lett.}\ }\textbf {\bibinfo {volume} {05}},\ \bibinfo {pages} {1287}
  (\bibinfo {year} {1998})}\BibitemShut {NoStop}%
\bibitem [{\citenamefont {Perdew}\ \emph {et~al.}(1996)\citenamefont {Perdew},
  \citenamefont {Burke},\ and\ \citenamefont {Ernzerhof}}]{Perdew1996}%
  \BibitemOpen
  \bibfield  {author} {\bibinfo {author} {\bibfnamefont {J.~P.}\ \bibnamefont
  {Perdew}}, \bibinfo {author} {\bibfnamefont {K.}~\bibnamefont {Burke}}, \
  and\ \bibinfo {author} {\bibfnamefont {M.}~\bibnamefont {Ernzerhof}},\ }\href
  {\doibase 10.1103/PhysRevLett.77.3865} {\bibfield  {journal} {\bibinfo
  {journal} {Phys. Rev. Lett.}\ }\textbf {\bibinfo {volume} {77}},\ \bibinfo
  {pages} {3865} (\bibinfo {year} {1996})}\BibitemShut {NoStop}%
\bibitem [{\citenamefont {Bl\"{o}chl}(1994)}]{Blochl1994}%
  \BibitemOpen
  \bibfield  {author} {\bibinfo {author} {\bibfnamefont {P.~E.}\ \bibnamefont
  {Bl\"{o}chl}},\ }\href {\doibase 10.1103/PhysRevB.50.17953} {\bibfield
  {journal} {\bibinfo  {journal} {Phys. Rev. B}\ }\textbf {\bibinfo {volume}
  {50}},\ \bibinfo {pages} {17953} (\bibinfo {year} {1994})}\BibitemShut
  {NoStop}%
\bibitem [{\citenamefont {Kresse}\ and\ \citenamefont
  {Hafner}(1993)}]{Kresse1993}%
  \BibitemOpen
  \bibfield  {author} {\bibinfo {author} {\bibfnamefont {G.}~\bibnamefont
  {Kresse}}\ and\ \bibinfo {author} {\bibfnamefont {J.}~\bibnamefont
  {Hafner}},\ }\href {\doibase 10.1103/PhysRevB.47.558} {\bibfield  {journal}
  {\bibinfo  {journal} {Phys. Rev. B}\ }\textbf {\bibinfo {volume} {47}},\
  \bibinfo {pages} {558} (\bibinfo {year} {1993})}\BibitemShut {NoStop}%
\bibitem [{\citenamefont {Kresse}\ and\ \citenamefont
  {Hafner}(1994)}]{Kresse1994}%
  \BibitemOpen
  \bibfield  {author} {\bibinfo {author} {\bibfnamefont {G.}~\bibnamefont
  {Kresse}}\ and\ \bibinfo {author} {\bibfnamefont {J.}~\bibnamefont
  {Hafner}},\ }\href {\doibase 10.1088/0953-8984/6/40/015} {\bibfield
  {journal} {\bibinfo  {journal} {J. Phys.: Condens. Matter}\ }\textbf
  {\bibinfo {volume} {6}},\ \bibinfo {pages} {8245} (\bibinfo {year}
  {1994})}\BibitemShut {NoStop}%
\bibitem [{\citenamefont {Anisimov}\ \emph {et~al.}(1997)\citenamefont
  {Anisimov}, \citenamefont {Aryasetiawan},\ and\ \citenamefont
  {Lichtenstein}}]{Anisimov1997}%
  \BibitemOpen
  \bibfield  {author} {\bibinfo {author} {\bibfnamefont {V.~I.}\ \bibnamefont
  {Anisimov}}, \bibinfo {author} {\bibfnamefont {F.}~\bibnamefont
  {Aryasetiawan}}, \ and\ \bibinfo {author} {\bibfnamefont {A.~I.}\
  \bibnamefont {Lichtenstein}},\ }\href {\doibase 10.1088/0953-8984/9/4/002}
  {\bibfield  {journal} {\bibinfo  {journal} {J. Phys.: Condens. Matter}\
  }\textbf {\bibinfo {volume} {9}},\ \bibinfo {pages} {767} (\bibinfo {year}
  {1997})}\BibitemShut {NoStop}%
\bibitem [{\citenamefont {Larson}\ \emph {et~al.}(2007)\citenamefont {Larson},
  \citenamefont {Lambrecht}, \citenamefont {Chantis},\ and\ \citenamefont {van
  Schilfgaarde}}]{Larson2007}%
  \BibitemOpen
  \bibfield  {author} {\bibinfo {author} {\bibfnamefont {P.}~\bibnamefont
  {Larson}}, \bibinfo {author} {\bibfnamefont {W.~R.~L.}\ \bibnamefont
  {Lambrecht}}, \bibinfo {author} {\bibfnamefont {A.}~\bibnamefont {Chantis}},
  \ and\ \bibinfo {author} {\bibfnamefont {M.}~\bibnamefont {van
  Schilfgaarde}},\ }\href {\doibase 10.1103/PhysRevB.75.045114} {\bibfield
  {journal} {\bibinfo  {journal} {Phys. Rev. B}\ }\textbf {\bibinfo {volume}
  {75}},\ \bibinfo {pages} {045114} (\bibinfo {year} {2007})}\BibitemShut
  {NoStop}%
\bibitem [{\citenamefont {Mallet}\ \emph {et~al.}(2007)\citenamefont {Mallet},
  \citenamefont {Varchon}, \citenamefont {Naud}, \citenamefont {Magaud},
  \citenamefont {Berger},\ and\ \citenamefont {Veuillen}}]{Mallet2007}%
  \BibitemOpen
  \bibfield  {author} {\bibinfo {author} {\bibfnamefont {P.}~\bibnamefont
  {Mallet}}, \bibinfo {author} {\bibfnamefont {F.}~\bibnamefont {Varchon}},
  \bibinfo {author} {\bibfnamefont {C.}~\bibnamefont {Naud}}, \bibinfo {author}
  {\bibfnamefont {L.}~\bibnamefont {Magaud}}, \bibinfo {author} {\bibfnamefont
  {C.}~\bibnamefont {Berger}}, \ and\ \bibinfo {author} {\bibfnamefont {J.-Y.}\
  \bibnamefont {Veuillen}},\ }\href {\doibase 10.1103/PhysRevB.76.041403}
  {\bibfield  {journal} {\bibinfo  {journal} {Phys. Rev. B.}\ }\textbf
  {\bibinfo {volume} {76}},\ \bibinfo {pages} {041403} (\bibinfo {year}
  {2007})}\BibitemShut {NoStop}%
\bibitem [{\citenamefont {Gr\r{a}näs}\ \emph {et~al.}(2012)\citenamefont
  {Gr\r{a}näs}, \citenamefont {Knudsen}, \citenamefont {Schröder},
  \citenamefont {Gerber}, \citenamefont {Busse}, \citenamefont {Arman},
  \citenamefont {Schulte}, \citenamefont {Andersen},\ and\ \citenamefont
  {Michely}}]{Granas2012}%
  \BibitemOpen
  \bibfield  {author} {\bibinfo {author} {\bibfnamefont {E.}~\bibnamefont
  {Gr\r{a}näs}}, \bibinfo {author} {\bibfnamefont {J.}~\bibnamefont {Knudsen}},
  \bibinfo {author} {\bibfnamefont {U.~A.}\ \bibnamefont {Schröder}}, \bibinfo
  {author} {\bibfnamefont {T.}~\bibnamefont {Gerber}}, \bibinfo {author}
  {\bibfnamefont {C.}~\bibnamefont {Busse}}, \bibinfo {author} {\bibfnamefont
  {M.~A.}\ \bibnamefont {Arman}}, \bibinfo {author} {\bibfnamefont
  {K.}~\bibnamefont {Schulte}}, \bibinfo {author} {\bibfnamefont {J.~N.}\
  \bibnamefont {Andersen}}, \ and\ \bibinfo {author} {\bibfnamefont
  {T.}~\bibnamefont {Michely}},\ }\href {\doibase 10.1021/nn303548z} {\bibfield
   {journal} {\bibinfo  {journal} {ACS Nano}\ }\textbf {\bibinfo {volume}
  {6}},\ \bibinfo {pages} {9951} (\bibinfo {year} {2012})}\BibitemShut
  {NoStop}%
\bibitem [{\citenamefont {Sutter}\ \emph {et~al.}(2010)\citenamefont {Sutter},
  \citenamefont {Sadowski},\ and\ \citenamefont {Sutter}}]{Sutter2010}%
  \BibitemOpen
  \bibfield  {author} {\bibinfo {author} {\bibfnamefont {P.}~\bibnamefont
  {Sutter}}, \bibinfo {author} {\bibfnamefont {J.~T.}\ \bibnamefont
  {Sadowski}}, \ and\ \bibinfo {author} {\bibfnamefont {E.~A.}\ \bibnamefont
  {Sutter}},\ }\href {\doibase 10.1021/ja102398n} {\bibfield  {journal}
  {\bibinfo  {journal} {J. Am. Chem. Soc.}\ }\textbf {\bibinfo {volume}
  {132}},\ \bibinfo {pages} {8175} (\bibinfo {year} {2010})}\BibitemShut
  {NoStop}%
\bibitem [{\citenamefont {Petrovi\'{c}}\ \emph {et~al.}(2013)\citenamefont
  {Petrovi\'{c}}, \citenamefont {{\v{S}rut Raki\'{c}}}, \citenamefont {Runte},
  \citenamefont {Busse}, \citenamefont {Sadowski}, \citenamefont {Lazi\'{c}},
  \citenamefont {Pletikosi\'{c}}, \citenamefont {Pan}, \citenamefont {Milun},
  \citenamefont {Pervan}, \citenamefont {Atodiresei}, \citenamefont {Brako},
  \citenamefont {Sokcevi\'{c}}, \citenamefont {Valla}, \citenamefont
  {Michely},\ and\ \citenamefont {Kralj}}]{Petrovic2013}%
  \BibitemOpen
  \bibfield  {author} {\bibinfo {author} {\bibfnamefont {M.}~\bibnamefont
  {Petrovi\'{c}}}, \bibinfo {author} {\bibfnamefont {I.}~\bibnamefont
  {{\v{S}rut Raki\'{c}}}}, \bibinfo {author} {\bibfnamefont {S.}~\bibnamefont
  {Runte}}, \bibinfo {author} {\bibfnamefont {C.}~\bibnamefont {Busse}},
  \bibinfo {author} {\bibfnamefont {J.~T.}\ \bibnamefont {Sadowski}}, \bibinfo
  {author} {\bibfnamefont {P.}~\bibnamefont {Lazi\'{c}}}, \bibinfo {author}
  {\bibfnamefont {I.}~\bibnamefont {Pletikosi\'{c}}}, \bibinfo {author}
  {\bibfnamefont {Z.-H.}\ \bibnamefont {Pan}}, \bibinfo {author} {\bibfnamefont
  {M.}~\bibnamefont {Milun}}, \bibinfo {author} {\bibfnamefont
  {P.}~\bibnamefont {Pervan}}, \bibinfo {author} {\bibfnamefont
  {N.}~\bibnamefont {Atodiresei}}, \bibinfo {author} {\bibfnamefont
  {R.}~\bibnamefont {Brako}}, \bibinfo {author} {\bibfnamefont
  {D.}~\bibnamefont {Sokcevi\'{c}}}, \bibinfo {author} {\bibfnamefont
  {T.}~\bibnamefont {Valla}}, \bibinfo {author} {\bibfnamefont
  {T.}~\bibnamefont {Michely}}, \ and\ \bibinfo {author} {\bibfnamefont
  {M.}~\bibnamefont {Kralj}},\ }\href {\doibase 10.1038/ncomms3772} {\bibfield
  {journal} {\bibinfo  {journal} {Nature Comm.}\ }\textbf {\bibinfo {volume}
  {4}},\ \bibinfo {pages} {2772} (\bibinfo {year} {2013})}\BibitemShut
  {NoStop}%
\bibitem [{\citenamefont {Markevich}\ \emph {et~al.}(2012)\citenamefont
  {Markevich}, \citenamefont {Jones}, \citenamefont {\"Oberg}, \citenamefont
  {Rayson}, \citenamefont {Goss},\ and\ \citenamefont
  {Briddon}}]{Markevich2012}%
  \BibitemOpen
  \bibfield  {author} {\bibinfo {author} {\bibfnamefont {A.}~\bibnamefont
  {Markevich}}, \bibinfo {author} {\bibfnamefont {R.}~\bibnamefont {Jones}},
  \bibinfo {author} {\bibfnamefont {S.}~\bibnamefont {\"Oberg}}, \bibinfo
  {author} {\bibfnamefont {M.~J.}\ \bibnamefont {Rayson}}, \bibinfo {author}
  {\bibfnamefont {J.~P.}\ \bibnamefont {Goss}}, \ and\ \bibinfo {author}
  {\bibfnamefont {P.~R.}\ \bibnamefont {Briddon}},\ }\href {\doibase
  10.1103/PhysRevB.86.045453} {\bibfield  {journal} {\bibinfo  {journal} {Phys.
  Rev. B}\ }\textbf {\bibinfo {volume} {86}},\ \bibinfo {pages} {045453}
  (\bibinfo {year} {2012})}\BibitemShut {NoStop}%
\bibitem [{\citenamefont {Kaloni}\ \emph {et~al.}(2012)\citenamefont {Kaloni},
  \citenamefont {Kahaly~Upadhyay}, \citenamefont {Cheng},\ and\ \citenamefont
  {Schwingenschl{\"o}gl}}]{Kaloni2012}%
  \BibitemOpen
  \bibfield  {author} {\bibinfo {author} {\bibfnamefont {T.~P.}\ \bibnamefont
  {Kaloni}}, \bibinfo {author} {\bibfnamefont {M.}~\bibnamefont
  {Kahaly~Upadhyay}}, \bibinfo {author} {\bibfnamefont {Y.~C.}\ \bibnamefont
  {Cheng}}, \ and\ \bibinfo {author} {\bibfnamefont {U.}~\bibnamefont
  {Schwingenschl{\"o}gl}},\ }\href {\doibase 10.1039/C2JM35127G} {\bibfield
  {journal} {\bibinfo  {journal} {J. Mater. Chem.}\ }\textbf {\bibinfo {volume}
  {22}},\ \bibinfo {pages} {23340} (\bibinfo {year} {2012})}\BibitemShut
  {NoStop}%
\bibitem [{\citenamefont {Xia}\ \emph {et~al.}(2012)\citenamefont {Xia},
  \citenamefont {Watcharinyanon}, \citenamefont {Zakharov}, \citenamefont
  {Yakimova}, \citenamefont {Hultman}, \citenamefont {Johansson},\ and\
  \citenamefont {Virojanadara}}]{Xia2012}%
  \BibitemOpen
  \bibfield  {author} {\bibinfo {author} {\bibfnamefont {C.}~\bibnamefont
  {Xia}}, \bibinfo {author} {\bibfnamefont {S.}~\bibnamefont {Watcharinyanon}},
  \bibinfo {author} {\bibfnamefont {A.~A.}\ \bibnamefont {Zakharov}}, \bibinfo
  {author} {\bibfnamefont {R.}~\bibnamefont {Yakimova}}, \bibinfo {author}
  {\bibfnamefont {L.}~\bibnamefont {Hultman}}, \bibinfo {author} {\bibfnamefont
  {L.~I.}\ \bibnamefont {Johansson}}, \ and\ \bibinfo {author} {\bibfnamefont
  {C.}~\bibnamefont {Virojanadara}},\ }\href {\doibase
  10.1103/PhysRevB.85.045418} {\bibfield  {journal} {\bibinfo  {journal} {Phys.
  Rev. B}\ }\textbf {\bibinfo {volume} {85}},\ \bibinfo {pages} {045418}
  (\bibinfo {year} {2012})}\BibitemShut {NoStop}%
\bibitem [{\citenamefont {Song}\ \emph {et~al.}(2013)\citenamefont {Song},
  \citenamefont {Ouyang},\ and\ \citenamefont {Medhekar}}]{Song2013}%
  \BibitemOpen
  \bibfield  {author} {\bibinfo {author} {\bibfnamefont {J.}~\bibnamefont
  {Song}}, \bibinfo {author} {\bibfnamefont {B.}~\bibnamefont {Ouyang}}, \ and\
  \bibinfo {author} {\bibfnamefont {N.~V.}\ \bibnamefont {Medhekar}},\ }\href
  {\doibase 10.1021/am403685w} {\bibfield  {journal} {\bibinfo  {journal} {ACS
  Appl. Mater. Inter.}\ }\textbf {\bibinfo {volume} {5}},\ \bibinfo {pages}
  {12968} (\bibinfo {year} {2013})}\BibitemShut {NoStop}%
\bibitem [{\citenamefont {Duong}\ \emph {et~al.}(2012)\citenamefont {Duong},
  \citenamefont {Han}, \citenamefont {Lee}, \citenamefont {Gunes},
  \citenamefont {Kim}, \citenamefont {Kim}, \citenamefont {Kim}, \citenamefont
  {Ta}, \citenamefont {So}, \citenamefont {Yoon}, \citenamefont {Chae},
  \citenamefont {Jo}, \citenamefont {Park}, \citenamefont {Chae}, \citenamefont
  {Lim}, \citenamefont {Choi},\ and\ \citenamefont {Lee}}]{Duong2012}%
  \BibitemOpen
  \bibfield  {author} {\bibinfo {author} {\bibfnamefont {D.~L.}\ \bibnamefont
  {Duong}}, \bibinfo {author} {\bibfnamefont {G.~H.}\ \bibnamefont {Han}},
  \bibinfo {author} {\bibfnamefont {S.~M.}\ \bibnamefont {Lee}}, \bibinfo
  {author} {\bibfnamefont {F.}~\bibnamefont {Gunes}}, \bibinfo {author}
  {\bibfnamefont {E.~S.}\ \bibnamefont {Kim}}, \bibinfo {author} {\bibfnamefont
  {S.~T.}\ \bibnamefont {Kim}}, \bibinfo {author} {\bibfnamefont
  {H.}~\bibnamefont {Kim}}, \bibinfo {author} {\bibfnamefont {Q.~H.}\
  \bibnamefont {Ta}}, \bibinfo {author} {\bibfnamefont {K.~P.}\ \bibnamefont
  {So}}, \bibinfo {author} {\bibfnamefont {S.~J.}\ \bibnamefont {Yoon}},
  \bibinfo {author} {\bibfnamefont {S.~J.}\ \bibnamefont {Chae}}, \bibinfo
  {author} {\bibfnamefont {Y.~W.}\ \bibnamefont {Jo}}, \bibinfo {author}
  {\bibfnamefont {M.~H.}\ \bibnamefont {Park}}, \bibinfo {author}
  {\bibfnamefont {S.~H.}\ \bibnamefont {Chae}}, \bibinfo {author}
  {\bibfnamefont {S.~C.}\ \bibnamefont {Lim}}, \bibinfo {author} {\bibfnamefont
  {J.~Y.}\ \bibnamefont {Choi}}, \ and\ \bibinfo {author} {\bibfnamefont
  {Y.~H.}\ \bibnamefont {Lee}},\ }\href {\doibase 10.1038/nature11562}
  {\bibfield  {journal} {\bibinfo  {journal} {Nature}\ }\textbf {\bibinfo
  {volume} {490}},\ \bibinfo {pages} {235} (\bibinfo {year}
  {2012})}\BibitemShut {NoStop}%
\bibitem [{\citenamefont {Sicot}\ \emph {et~al.}(2012)\citenamefont {Sicot},
  \citenamefont {Leicht}, \citenamefont {Zusan}, \citenamefont {Bouvron},
  \citenamefont {Zander}, \citenamefont {Weser}, \citenamefont {Dedkov},
  \citenamefont {Horn},\ and\ \citenamefont {Fonin}}]{Sicot2012}%
  \BibitemOpen
  \bibfield  {author} {\bibinfo {author} {\bibfnamefont {M.}~\bibnamefont
  {Sicot}}, \bibinfo {author} {\bibfnamefont {P.}~\bibnamefont {Leicht}},
  \bibinfo {author} {\bibfnamefont {A.}~\bibnamefont {Zusan}}, \bibinfo
  {author} {\bibfnamefont {S.}~\bibnamefont {Bouvron}}, \bibinfo {author}
  {\bibfnamefont {O.}~\bibnamefont {Zander}}, \bibinfo {author} {\bibfnamefont
  {M.}~\bibnamefont {Weser}}, \bibinfo {author} {\bibfnamefont
  {Y.}~\bibnamefont {Dedkov}}, \bibinfo {author} {\bibfnamefont
  {K.}~\bibnamefont {Horn}}, \ and\ \bibinfo {author} {\bibfnamefont
  {M.}~\bibnamefont {Fonin}},\ }\href {\doibase 10.1021/nn203169j} {\bibfield
  {journal} {\bibinfo  {journal} {ACS Nano}\ }\textbf {\bibinfo {volume}
  {114}},\ \bibinfo {pages} {151} (\bibinfo {year} {2012})}\BibitemShut
  {NoStop}%
\bibitem [{\citenamefont {Boukhvalov}\ and\ \citenamefont
  {Katsnelson}(2009)}]{Boukhvalov2009}%
  \BibitemOpen
  \bibfield  {author} {\bibinfo {author} {\bibfnamefont {D.~W.}\ \bibnamefont
  {Boukhvalov}}\ and\ \bibinfo {author} {\bibfnamefont {M.~I.}\ \bibnamefont
  {Katsnelson}},\ }\href {\doibase 10.1063/1.3160551} {\bibfield  {journal}
  {\bibinfo  {journal} {Appl. Phys. Lett.}\ }\textbf {\bibinfo {volume} {95}},\
  \bibinfo {pages} {023109} (\bibinfo {year} {2009})}\BibitemShut {NoStop}%
\bibitem [{\citenamefont {Virojanadara}\ \emph {et~al.}(2010)\citenamefont
  {Virojanadara}, \citenamefont {Watcharinyanon}, \citenamefont {Zakharov},\
  and\ \citenamefont {Johansson}}]{Virojanadara2010}%
  \BibitemOpen
  \bibfield  {author} {\bibinfo {author} {\bibfnamefont {C.}~\bibnamefont
  {Virojanadara}}, \bibinfo {author} {\bibfnamefont {S.}~\bibnamefont
  {Watcharinyanon}}, \bibinfo {author} {\bibfnamefont {A.~A.}\ \bibnamefont
  {Zakharov}}, \ and\ \bibinfo {author} {\bibfnamefont {L.~I.}\ \bibnamefont
  {Johansson}},\ }\href {\doibase 10.1103/PhysRevB.82.205402} {\bibfield
  {journal} {\bibinfo  {journal} {Phys. Rev. B}\ }\textbf {\bibinfo {volume}
  {82}},\ \bibinfo {pages} {205402} (\bibinfo {year} {2010})}\BibitemShut
  {NoStop}%
\bibitem [{Sup()}]{Supplement}%
  \BibitemOpen
  \href@noop {} {}\bibinfo {note} {See Supplemental Material for more
  information about the magnetic properties, the intercalation process, the
  ARPES geometry, and Raman spectroscopy.}\BibitemShut {Stop}%
\bibitem [{\citenamefont {Coraux}\ \emph {et~al.}(2008)\citenamefont {Coraux},
  \citenamefont {N'Diaye}, \citenamefont {Busse},\ and\ \citenamefont
  {Michely}}]{Coraux2008}%
  \BibitemOpen
  \bibfield  {author} {\bibinfo {author} {\bibfnamefont {J.}~\bibnamefont
  {Coraux}}, \bibinfo {author} {\bibfnamefont {A.}~\bibnamefont {N'Diaye}},
  \bibinfo {author} {\bibfnamefont {C.}~\bibnamefont {Busse}}, \ and\ \bibinfo
  {author} {\bibfnamefont {T.}~\bibnamefont {Michely}},\ }\href {\doibase
  10.1021/nl0728874} {\bibfield  {journal} {\bibinfo  {journal} {Nano Lett.}\
  }\textbf {\bibinfo {volume} {8}},\ \bibinfo {pages} {565} (\bibinfo {year}
  {2008})}\BibitemShut {NoStop}%
\bibitem [{\citenamefont {Lehtinen}\ \emph {et~al.}(2013)\citenamefont
  {Lehtinen}, \citenamefont {Kurasch}, \citenamefont {Krasheninnikov},\ and\
  \citenamefont {Kaiser}}]{Lehtinen2013}%
  \BibitemOpen
  \bibfield  {author} {\bibinfo {author} {\bibfnamefont {O.}~\bibnamefont
  {Lehtinen}}, \bibinfo {author} {\bibfnamefont {S.}~\bibnamefont {Kurasch}},
  \bibinfo {author} {\bibfnamefont {A.~V.}\ \bibnamefont {Krasheninnikov}}, \
  and\ \bibinfo {author} {\bibfnamefont {U.}~\bibnamefont {Kaiser}},\ }\href
  {\doibase 10.1038/ncomms3098} {\bibfield  {journal} {\bibinfo  {journal}
  {Nature Comm.}\ }\textbf {\bibinfo {volume} {4}},\ \bibinfo {pages} {2098}
  (\bibinfo {year} {2013})}\BibitemShut {NoStop}%
\bibitem [{\citenamefont {Thole}\ \emph {et~al.}(1992)\citenamefont {Thole},
  \citenamefont {Carra}, \citenamefont {Sette},\ and\ \citenamefont {van~der
  Laan}}]{Thole1992}%
  \BibitemOpen
  \bibfield  {author} {\bibinfo {author} {\bibfnamefont {B.~T.}\ \bibnamefont
  {Thole}}, \bibinfo {author} {\bibfnamefont {P.}~\bibnamefont {Carra}},
  \bibinfo {author} {\bibfnamefont {F.}~\bibnamefont {Sette}}, \ and\ \bibinfo
  {author} {\bibfnamefont {G.}~\bibnamefont {van~der Laan}},\ }\href {\doibase
  10.1103/PhysRevLett.68.1943} {\bibfield  {journal} {\bibinfo  {journal}
  {Phys. Rev. Lett.}\ }\textbf {\bibinfo {volume} {68}},\ \bibinfo {pages}
  {1943} (\bibinfo {year} {1992})}\BibitemShut {NoStop}%
\bibitem [{\citenamefont {Carra}\ \emph {et~al.}(1993)\citenamefont {Carra},
  \citenamefont {Thole}, \citenamefont {Altarelli},\ and\ \citenamefont
  {Wang}}]{Carra1993}%
  \BibitemOpen
  \bibfield  {author} {\bibinfo {author} {\bibfnamefont {P.}~\bibnamefont
  {Carra}}, \bibinfo {author} {\bibfnamefont {B.~T.}\ \bibnamefont {Thole}},
  \bibinfo {author} {\bibfnamefont {M.}~\bibnamefont {Altarelli}}, \ and\
  \bibinfo {author} {\bibfnamefont {X.}~\bibnamefont {Wang}},\ }\href {\doibase
  10.1103/PhysRevLett.70.694} {\bibfield  {journal} {\bibinfo  {journal} {Phys.
  Rev. Lett.}\ }\textbf {\bibinfo {volume} {70}},\ \bibinfo {pages} {694}
  (\bibinfo {year} {1993})}\BibitemShut {NoStop}%
\bibitem [{\citenamefont {Wu}\ and\ \citenamefont {Freeman}(1994)}]{Wu1994}%
  \BibitemOpen
  \bibfield  {author} {\bibinfo {author} {\bibfnamefont {R.}~\bibnamefont
  {Wu}}\ and\ \bibinfo {author} {\bibfnamefont {A.~J.}\ \bibnamefont
  {Freeman}},\ }\href {\doibase 10.1103/PhysRevLett.73.1994} {\bibfield
  {journal} {\bibinfo  {journal} {Phys. Rev. Lett.}\ }\textbf {\bibinfo
  {volume} {73}},\ \bibinfo {pages} {1994} (\bibinfo {year}
  {1994})}\BibitemShut {NoStop}%
\bibitem [{\citenamefont {Crocombette}\ \emph {et~al.}(1996)\citenamefont
  {Crocombette}, \citenamefont {Thole},\ and\ \citenamefont
  {Jollet}}]{Crocombette1996}%
  \BibitemOpen
  \bibfield  {author} {\bibinfo {author} {\bibfnamefont {J.~P.}\ \bibnamefont
  {Crocombette}}, \bibinfo {author} {\bibfnamefont {B.~T.}\ \bibnamefont
  {Thole}}, \ and\ \bibinfo {author} {\bibfnamefont {F.}~\bibnamefont
  {Jollet}},\ }\href {\doibase 10.1088/0953-8984/8/22/013} {\bibfield
  {journal} {\bibinfo  {journal} {J. Phys.: Condens. Matter}\ }\textbf
  {\bibinfo {volume} {8}},\ \bibinfo {pages} {4095} (\bibinfo {year}
  {1996})}\BibitemShut {NoStop}%
\bibitem [{\citenamefont {Schill\'e}\ \emph {et~al.}(1993)\citenamefont
  {Schill\'e}, \citenamefont {Kappler}, \citenamefont {Sainctavit},
  \citenamefont {Cartier~dit Moulin}, \citenamefont {Brouder},\ and\
  \citenamefont {Krill}}]{Schille1993}%
  \BibitemOpen
  \bibfield  {author} {\bibinfo {author} {\bibfnamefont {J.~P.}\ \bibnamefont
  {Schill\'e}}, \bibinfo {author} {\bibfnamefont {J.~P.}\ \bibnamefont
  {Kappler}}, \bibinfo {author} {\bibfnamefont {P.}~\bibnamefont {Sainctavit}},
  \bibinfo {author} {\bibfnamefont {C.}~\bibnamefont {Cartier~dit Moulin}},
  \bibinfo {author} {\bibfnamefont {C.}~\bibnamefont {Brouder}}, \ and\
  \bibinfo {author} {\bibfnamefont {G.}~\bibnamefont {Krill}},\ }\href
  {\doibase 10.1103/PhysRevB.48.9491} {\bibfield  {journal} {\bibinfo
  {journal} {Phys. Rev. B}\ }\textbf {\bibinfo {volume} {48}},\ \bibinfo
  {pages} {9491} (\bibinfo {year} {1993})}\BibitemShut {NoStop}%
\bibitem [{\citenamefont {Thole}\ \emph {et~al.}(1985)\citenamefont {Thole},
  \citenamefont {van~der Laan}, \citenamefont {Fuggle}, \citenamefont
  {Sawatzky}, \citenamefont {Karnatak},\ and\ \citenamefont
  {Esteva}}]{Thole1985}%
  \BibitemOpen
  \bibfield  {author} {\bibinfo {author} {\bibfnamefont {B.~T.}\ \bibnamefont
  {Thole}}, \bibinfo {author} {\bibfnamefont {G.}~\bibnamefont {van~der Laan}},
  \bibinfo {author} {\bibfnamefont {J.~C.}\ \bibnamefont {Fuggle}}, \bibinfo
  {author} {\bibfnamefont {G.~A.}\ \bibnamefont {Sawatzky}}, \bibinfo {author}
  {\bibfnamefont {R.~C.}\ \bibnamefont {Karnatak}}, \ and\ \bibinfo {author}
  {\bibfnamefont {J.-M.}\ \bibnamefont {Esteva}},\ }\href {\doibase
  10.1103/PhysRevB.32.5107} {\bibfield  {journal} {\bibinfo  {journal} {Phys.
  Rev. B}\ }\textbf {\bibinfo {volume} {32}},\ \bibinfo {pages} {5107}
  (\bibinfo {year} {1985})}\BibitemShut {NoStop}%
\bibitem [{\citenamefont {F\"{o}rster}\ \emph {et~al.}(2011)\citenamefont
  {F\"{o}rster}, \citenamefont {Klinkhammer}, \citenamefont {Busse},
  \citenamefont {Altendorf}, \citenamefont {Michely}, \citenamefont {Hu},
  \citenamefont {Chin}, \citenamefont {Tjeng}, \citenamefont {Coraux},\ and\
  \citenamefont {Bourgault}}]{Forster2011}%
  \BibitemOpen
  \bibfield  {author} {\bibinfo {author} {\bibfnamefont {D.~F.}\ \bibnamefont
  {F\"{o}rster}}, \bibinfo {author} {\bibfnamefont {J.}~\bibnamefont
  {Klinkhammer}}, \bibinfo {author} {\bibfnamefont {C.}~\bibnamefont {Busse}},
  \bibinfo {author} {\bibfnamefont {S.~G.}\ \bibnamefont {Altendorf}}, \bibinfo
  {author} {\bibfnamefont {T.}~\bibnamefont {Michely}}, \bibinfo {author}
  {\bibfnamefont {Z.}~\bibnamefont {Hu}}, \bibinfo {author} {\bibfnamefont
  {Y.-Y.}\ \bibnamefont {Chin}}, \bibinfo {author} {\bibfnamefont {L.~H.}\
  \bibnamefont {Tjeng}}, \bibinfo {author} {\bibfnamefont {J.}~\bibnamefont
  {Coraux}}, \ and\ \bibinfo {author} {\bibfnamefont {D.}~\bibnamefont
  {Bourgault}},\ }\href {\doibase 10.1103/PhysRevB.83.045424} {\bibfield
  {journal} {\bibinfo  {journal} {Phys. Rev. B}\ }\textbf {\bibinfo {volume}
  {83}},\ \bibinfo {pages} {045424} (\bibinfo {year} {2011})}\BibitemShut
  {NoStop}%
\bibitem [{\citenamefont {Mermin}\ and\ \citenamefont
  {Wagner}(1966)}]{Mermin1966}%
  \BibitemOpen
  \bibfield  {author} {\bibinfo {author} {\bibfnamefont {N.~D.}\ \bibnamefont
  {Mermin}}\ and\ \bibinfo {author} {\bibfnamefont {H.}~\bibnamefont
  {Wagner}},\ }\href {\doibase 10.1103/PhysRevLett.17.1133} {\bibfield
  {journal} {\bibinfo  {journal} {Phys. Rev. Lett.}\ }\textbf {\bibinfo
  {volume} {17}},\ \bibinfo {pages} {1133} (\bibinfo {year}
  {1966})}\BibitemShut {NoStop}%
\bibitem [{\citenamefont {Grechnev}\ \emph {et~al.}(2005)\citenamefont
  {Grechnev}, \citenamefont {Irkhin}, \citenamefont {Katsnelson},\ and\
  \citenamefont {Eriksson}}]{Grechnev2005}%
  \BibitemOpen
  \bibfield  {author} {\bibinfo {author} {\bibfnamefont {A.}~\bibnamefont
  {Grechnev}}, \bibinfo {author} {\bibfnamefont {V.~Y.}\ \bibnamefont
  {Irkhin}}, \bibinfo {author} {\bibfnamefont {M.~I.}\ \bibnamefont
  {Katsnelson}}, \ and\ \bibinfo {author} {\bibfnamefont {O.}~\bibnamefont
  {Eriksson}},\ }\href {\doibase 10.1103/PhysRevB.71.024427} {\bibfield
  {journal} {\bibinfo  {journal} {Phys. Rev. B}\ }\textbf {\bibinfo {volume}
  {71}},\ \bibinfo {pages} {024427} (\bibinfo {year} {2005})}\BibitemShut
  {NoStop}%
\bibitem [{\citenamefont {Johnson}\ \emph {et~al.}(1996)\citenamefont
  {Johnson}, \citenamefont {Bloemen}, \citenamefont {den Broeder},\ and\
  \citenamefont {de~Vries}}]{Johnson1996}%
  \BibitemOpen
  \bibfield  {author} {\bibinfo {author} {\bibfnamefont {M.~T.}\ \bibnamefont
  {Johnson}}, \bibinfo {author} {\bibfnamefont {P.~J.~H.}\ \bibnamefont
  {Bloemen}}, \bibinfo {author} {\bibfnamefont {F.~J.~A.}\ \bibnamefont {den
  Broeder}}, \ and\ \bibinfo {author} {\bibfnamefont {J.~J.}\ \bibnamefont
  {de~Vries}},\ }\href {\doibase 10.1088/0034-4885/59/11/002} {\bibfield
  {journal} {\bibinfo  {journal} {Rep. Prog. Phys.}\ }\textbf {\bibinfo
  {volume} {59}},\ \bibinfo {pages} {1409} (\bibinfo {year}
  {1996})}\BibitemShut {NoStop}%
\bibitem [{\citenamefont {Daalderop}\ \emph {et~al.}(1990)\citenamefont
  {Daalderop}, \citenamefont {Kelly},\ and\ \citenamefont
  {Schuurmans}}]{Daalderop1990}%
  \BibitemOpen
  \bibfield  {author} {\bibinfo {author} {\bibfnamefont {G.~H.~O.}\
  \bibnamefont {Daalderop}}, \bibinfo {author} {\bibfnamefont {P.~J.}\
  \bibnamefont {Kelly}}, \ and\ \bibinfo {author} {\bibfnamefont {M.~F.~H.}\
  \bibnamefont {Schuurmans}},\ }\href {\doibase 10.1103/PhysRevB.41.11919}
  {\bibfield  {journal} {\bibinfo  {journal} {Phys. Rev. B}\ }\textbf {\bibinfo
  {volume} {41}},\ \bibinfo {pages} {11919} (\bibinfo {year}
  {1990})}\BibitemShut {NoStop}%
\bibitem [{\citenamefont {Barmak}\ \emph {et~al.}(2005)\citenamefont {Barmak},
  \citenamefont {Kim}, \citenamefont {Lewis}, \citenamefont {Coffey},
  \citenamefont {Toney}, \citenamefont {Kellock},\ and\ \citenamefont
  {Thiele}}]{Barmak2005}%
  \BibitemOpen
  \bibfield  {author} {\bibinfo {author} {\bibfnamefont {K.}~\bibnamefont
  {Barmak}}, \bibinfo {author} {\bibfnamefont {J.}~\bibnamefont {Kim}},
  \bibinfo {author} {\bibfnamefont {L.~H.}\ \bibnamefont {Lewis}}, \bibinfo
  {author} {\bibfnamefont {K.~R.}\ \bibnamefont {Coffey}}, \bibinfo {author}
  {\bibfnamefont {M.~F.}\ \bibnamefont {Toney}}, \bibinfo {author}
  {\bibfnamefont {A.~J.}\ \bibnamefont {Kellock}}, \ and\ \bibinfo {author}
  {\bibfnamefont {J.-U.}\ \bibnamefont {Thiele}},\ }\href {\doibase
  10.1063/1.1991968} {\bibfield  {journal} {\bibinfo  {journal} {J. Appl.
  Phys.}\ }\textbf {\bibinfo {volume} {98}},\ \bibinfo {pages} {033904}
  (\bibinfo {year} {2005})}\BibitemShut {NoStop}%
\bibitem [{\citenamefont {Deutscher}\ and\ \citenamefont
  {Everts}(1993)}]{Deutscher1993}%
  \BibitemOpen
  \bibfield  {author} {\bibinfo {author} {\bibfnamefont {R.}~\bibnamefont
  {Deutscher}}\ and\ \bibinfo {author} {\bibfnamefont {H.~U.}\ \bibnamefont
  {Everts}},\ }\href {\doibase 10.1007/BF01308811} {\bibfield  {journal}
  {\bibinfo  {journal} {Z. Phys. B: Condens. Matter}\ }\textbf {\bibinfo
  {volume} {93}},\ \bibinfo {pages} {77} (\bibinfo {year} {1993})}\BibitemShut
  {NoStop}%
\bibitem [{\citenamefont {Suematsu}\ \emph {et~al.}(1981)\citenamefont
  {Suematsu}, \citenamefont {Ohmatsu},\ and\ \citenamefont
  {Yoshizaki}}]{Suematsu1981a}%
  \BibitemOpen
  \bibfield  {author} {\bibinfo {author} {\bibfnamefont {H.}~\bibnamefont
  {Suematsu}}, \bibinfo {author} {\bibfnamefont {K.}~\bibnamefont {Ohmatsu}}, \
  and\ \bibinfo {author} {\bibfnamefont {R.}~\bibnamefont {Yoshizaki}},\ }\href
  {\doibase 10.1016/0038-1098(81)90966-2} {\bibfield  {journal} {\bibinfo
  {journal} {Solid State Commun.}\ }\textbf {\bibinfo {volume} {38}},\ \bibinfo
  {pages} {1103} (\bibinfo {year} {1981})}\BibitemShut {NoStop}%
\bibitem [{\citenamefont {Mucha-Kruczy\'{n}ski}\ \emph
  {et~al.}(2008)\citenamefont {Mucha-Kruczy\'{n}ski}, \citenamefont
  {Tsyplyatyev}, \citenamefont {Grishin}, \citenamefont {McCann}, \citenamefont
  {Fal'ko}, \citenamefont {Bostwick},\ and\ \citenamefont
  {Rotenberg}}]{Mucha2008}%
  \BibitemOpen
  \bibfield  {author} {\bibinfo {author} {\bibfnamefont {M.}~\bibnamefont
  {Mucha-Kruczy\'{n}ski}}, \bibinfo {author} {\bibfnamefont {O.}~\bibnamefont
  {Tsyplyatyev}}, \bibinfo {author} {\bibfnamefont {A.}~\bibnamefont
  {Grishin}}, \bibinfo {author} {\bibfnamefont {E.}~\bibnamefont {McCann}},
  \bibinfo {author} {\bibfnamefont {V.~I.}\ \bibnamefont {Fal'ko}}, \bibinfo
  {author} {\bibfnamefont {A.}~\bibnamefont {Bostwick}}, \ and\ \bibinfo
  {author} {\bibfnamefont {E.}~\bibnamefont {Rotenberg}},\ }\href {\doibase
  10.1103/PhysRevB.77.195403} {\bibfield  {journal} {\bibinfo  {journal} {Phys.
  Rev. B}\ }\textbf {\bibinfo {volume} {77}},\ \bibinfo {pages} {195403}
  (\bibinfo {year} {2008})}\BibitemShut {NoStop}%
\bibitem [{\citenamefont {Pletikosi{\'c}}\ \emph {et~al.}(2009)\citenamefont
  {Pletikosi{\'c}}, \citenamefont {Kralj}, \citenamefont {Pervan},
  \citenamefont {Brako}, \citenamefont {Coraux}, \citenamefont {N'Diaye},
  \citenamefont {Busse},\ and\ \citenamefont {Michely}}]{Pletikosic2009}%
  \BibitemOpen
  \bibfield  {author} {\bibinfo {author} {\bibfnamefont {I.}~\bibnamefont
  {Pletikosi{\'c}}}, \bibinfo {author} {\bibfnamefont {M.}~\bibnamefont
  {Kralj}}, \bibinfo {author} {\bibfnamefont {P.}~\bibnamefont {Pervan}},
  \bibinfo {author} {\bibfnamefont {R.}~\bibnamefont {Brako}}, \bibinfo
  {author} {\bibfnamefont {J.}~\bibnamefont {Coraux}}, \bibinfo {author}
  {\bibfnamefont {A.~T.}\ \bibnamefont {N'Diaye}}, \bibinfo {author}
  {\bibfnamefont {C.}~\bibnamefont {Busse}}, \ and\ \bibinfo {author}
  {\bibfnamefont {T.}~\bibnamefont {Michely}},\ }\href {\doibase
  10.1103/PhysRevLett.102.056808} {\bibfield  {journal} {\bibinfo  {journal}
  {Phys. Rev. Lett.}\ }\textbf {\bibinfo {volume} {102}},\ \bibinfo {pages}
  {056808} (\bibinfo {year} {2009})}\BibitemShut {NoStop}%
\bibitem [{\citenamefont {Castro~Neto}\ \emph {et~al.}(2009)\citenamefont
  {Castro~Neto}, \citenamefont {Guinea}, \citenamefont {Peres}, \citenamefont
  {Novoselov},\ and\ \citenamefont {Geim}}]{CastroNeto2009}%
  \BibitemOpen
  \bibfield  {author} {\bibinfo {author} {\bibfnamefont {A.~H.}\ \bibnamefont
  {Castro~Neto}}, \bibinfo {author} {\bibfnamefont {F.}~\bibnamefont {Guinea}},
  \bibinfo {author} {\bibfnamefont {N.~M.~R.}\ \bibnamefont {Peres}}, \bibinfo
  {author} {\bibfnamefont {K.~S.}\ \bibnamefont {Novoselov}}, \ and\ \bibinfo
  {author} {\bibfnamefont {A.~K.}\ \bibnamefont {Geim}},\ }\href {\doibase
  10.1103/RevModPhys.81.109} {\bibfield  {journal} {\bibinfo  {journal} {Rev.
  Mod. Phys.}\ }\textbf {\bibinfo {volume} {81}},\ \bibinfo {pages} {109}
  (\bibinfo {year} {2009})}\BibitemShut {NoStop}%
\bibitem [{\citenamefont {Kralj}\ \emph {et~al.}(2011)\citenamefont {Kralj},
  \citenamefont {Pletikosi\ifmmode~\acute{c}\else \'{c}\fi{}}, \citenamefont
  {Petrovi\ifmmode~\acute{c}\else \'{c}\fi{}}, \citenamefont {Pervan},
  \citenamefont {Milun}, \citenamefont {N'Diaye}, \citenamefont {Busse},
  \citenamefont {Michely}, \citenamefont {Fujii},\ and\ \citenamefont
  {Vobornik}}]{Kralj2011}%
  \BibitemOpen
  \bibfield  {author} {\bibinfo {author} {\bibfnamefont {M.}~\bibnamefont
  {Kralj}}, \bibinfo {author} {\bibfnamefont {I.}~\bibnamefont
  {Pletikosi\ifmmode~\acute{c}\else \'{c}\fi{}}}, \bibinfo {author}
  {\bibfnamefont {M.}~\bibnamefont {Petrovi\ifmmode~\acute{c}\else
  \'{c}\fi{}}}, \bibinfo {author} {\bibfnamefont {P.}~\bibnamefont {Pervan}},
  \bibinfo {author} {\bibfnamefont {M.}~\bibnamefont {Milun}}, \bibinfo
  {author} {\bibfnamefont {A.~T.}\ \bibnamefont {N'Diaye}}, \bibinfo {author}
  {\bibfnamefont {C.}~\bibnamefont {Busse}}, \bibinfo {author} {\bibfnamefont
  {T.}~\bibnamefont {Michely}}, \bibinfo {author} {\bibfnamefont
  {J.}~\bibnamefont {Fujii}}, \ and\ \bibinfo {author} {\bibfnamefont
  {I.}~\bibnamefont {Vobornik}},\ }\href {\doibase 10.1103/PhysRevB.84.075427}
  {\bibfield  {journal} {\bibinfo  {journal} {Phys. Rev. B}\ }\textbf {\bibinfo
  {volume} {84}},\ \bibinfo {pages} {075427} (\bibinfo {year}
  {2011})}\BibitemShut {NoStop}%
\bibitem [{\citenamefont {Bostwick}\ \emph {et~al.}(2007)\citenamefont
  {Bostwick}, \citenamefont {Ohta}, \citenamefont {McChesney}, \citenamefont
  {Seyller}, \citenamefont {Horn},\ and\ \citenamefont
  {Rotenberg}}]{Bostwick2007}%
  \BibitemOpen
  \bibfield  {author} {\bibinfo {author} {\bibfnamefont {A.}~\bibnamefont
  {Bostwick}}, \bibinfo {author} {\bibfnamefont {T.}~\bibnamefont {Ohta}},
  \bibinfo {author} {\bibfnamefont {J.~L.}\ \bibnamefont {McChesney}}, \bibinfo
  {author} {\bibfnamefont {T.}~\bibnamefont {Seyller}}, \bibinfo {author}
  {\bibfnamefont {K.}~\bibnamefont {Horn}}, \ and\ \bibinfo {author}
  {\bibfnamefont {E.}~\bibnamefont {Rotenberg}},\ }\href {\doibase
  10.1016/j.ssc.2007.04.034} {\bibfield  {journal} {\bibinfo  {journal} {Solid
  State Commun.}\ }\textbf {\bibinfo {volume} {143}},\ \bibinfo {pages} {63}
  (\bibinfo {year} {2007})}\BibitemShut {NoStop}%
\bibitem [{\citenamefont {Tse}\ and\ \citenamefont
  {Das~Sarma}(2007)}]{Tse2007}%
  \BibitemOpen
  \bibfield  {author} {\bibinfo {author} {\bibfnamefont {W.-K.}\ \bibnamefont
  {Tse}}\ and\ \bibinfo {author} {\bibfnamefont {S.}~\bibnamefont
  {Das~Sarma}},\ }\href {\doibase 10.1103/PhysRevLett.99.236802} {\bibfield
  {journal} {\bibinfo  {journal} {Phys. Rev. Lett.}\ }\textbf {\bibinfo
  {volume} {99}},\ \bibinfo {pages} {236802} (\bibinfo {year}
  {2007})}\BibitemShut {NoStop}%
\bibitem [{\citenamefont {Starodub}\ \emph {et~al.}(2011)\citenamefont
  {Starodub}, \citenamefont {Bostwick}, \citenamefont {Moreschini},
  \citenamefont {Nie}, \citenamefont {{El Gabaly}}, \citenamefont {McCarthy},\
  and\ \citenamefont {Rotenberg}}]{Starodub2011}%
  \BibitemOpen
  \bibfield  {author} {\bibinfo {author} {\bibfnamefont {E.}~\bibnamefont
  {Starodub}}, \bibinfo {author} {\bibfnamefont {A.}~\bibnamefont {Bostwick}},
  \bibinfo {author} {\bibfnamefont {L.}~\bibnamefont {Moreschini}}, \bibinfo
  {author} {\bibfnamefont {S.}~\bibnamefont {Nie}}, \bibinfo {author}
  {\bibfnamefont {F.}~\bibnamefont {{El Gabaly}}}, \bibinfo {author}
  {\bibfnamefont {K.~F.}\ \bibnamefont {McCarthy}}, \ and\ \bibinfo {author}
  {\bibfnamefont {E.}~\bibnamefont {Rotenberg}},\ }\href {\doibase
  10.1103/PhysRevB.83.125428} {\bibfield  {journal} {\bibinfo  {journal} {Phys.
  Rev. B}\ }\textbf {\bibinfo {volume} {83}},\ \bibinfo {pages} {125428}
  (\bibinfo {year} {2011})}\BibitemShut {NoStop}%
\bibitem [{\citenamefont {Ferrari}\ and\ \citenamefont
  {Basko}(2013)}]{Ferrari13}%
  \BibitemOpen
  \bibfield  {author} {\bibinfo {author} {\bibfnamefont {A.~C.}\ \bibnamefont
  {Ferrari}}\ and\ \bibinfo {author} {\bibfnamefont {D.~M.}\ \bibnamefont
  {Basko}},\ }\href {\doibase 10.1038/nnano.2013.46} {\bibfield  {journal}
  {\bibinfo  {journal} {Nature Nanotech.}\ }\textbf {\bibinfo {volume} {8}},\
  \bibinfo {pages} {235} (\bibinfo {year} {2013})}\BibitemShut {NoStop}%
\bibitem [{\citenamefont {Das}\ \emph {et~al.}(2008)\citenamefont {Das},
  \citenamefont {Pisana}, \citenamefont {Chakraborty}, \citenamefont
  {Piscanec}, \citenamefont {Saha}, \citenamefont {Waghmare}, \citenamefont
  {Novoselov}, \citenamefont {Krishnamurthy}, \citenamefont {Geim},
  \citenamefont {Ferrari},\ and\ \citenamefont {Sood}}]{Das08}%
  \BibitemOpen
  \bibfield  {author} {\bibinfo {author} {\bibfnamefont {A.}~\bibnamefont
  {Das}}, \bibinfo {author} {\bibfnamefont {S.}~\bibnamefont {Pisana}},
  \bibinfo {author} {\bibfnamefont {B.}~\bibnamefont {Chakraborty}}, \bibinfo
  {author} {\bibfnamefont {S.}~\bibnamefont {Piscanec}}, \bibinfo {author}
  {\bibfnamefont {S.~K.}\ \bibnamefont {Saha}}, \bibinfo {author}
  {\bibfnamefont {U.~V.}\ \bibnamefont {Waghmare}}, \bibinfo {author}
  {\bibfnamefont {K.~S.}\ \bibnamefont {Novoselov}}, \bibinfo {author}
  {\bibfnamefont {H.~R.}\ \bibnamefont {Krishnamurthy}}, \bibinfo {author}
  {\bibfnamefont {A.~K.}\ \bibnamefont {Geim}}, \bibinfo {author}
  {\bibfnamefont {A.~C.}\ \bibnamefont {Ferrari}}, \ and\ \bibinfo {author}
  {\bibfnamefont {A.~K.}\ \bibnamefont {Sood}},\ }\href {\doibase
  10.1038/nnano.2008.67} {\bibfield  {journal} {\bibinfo  {journal} {Nature
  Nanotech.}\ }\textbf {\bibinfo {volume} {3}},\ \bibinfo {pages} {210}
  (\bibinfo {year} {2008})}\BibitemShut {NoStop}%
\bibitem [{\citenamefont {Kalbac}\ \emph {et~al.}(2010)\citenamefont {Kalbac},
  \citenamefont {Reina-Cecco}, \citenamefont {Farhat}, \citenamefont {Kong},
  \citenamefont {Kavan},\ and\ \citenamefont {Dresselhaus}}]{Kalbac10}%
  \BibitemOpen
  \bibfield  {author} {\bibinfo {author} {\bibfnamefont {M.}~\bibnamefont
  {Kalbac}}, \bibinfo {author} {\bibfnamefont {A.}~\bibnamefont {Reina-Cecco}},
  \bibinfo {author} {\bibfnamefont {H.}~\bibnamefont {Farhat}}, \bibinfo
  {author} {\bibfnamefont {J.}~\bibnamefont {Kong}}, \bibinfo {author}
  {\bibfnamefont {L.}~\bibnamefont {Kavan}}, \ and\ \bibinfo {author}
  {\bibfnamefont {M.~S.}\ \bibnamefont {Dresselhaus}},\ }\href {\doibase
  10.1021/nn1010914} {\bibfield  {journal} {\bibinfo  {journal} {ACS Nano}\
  }\textbf {\bibinfo {volume} {4}},\ \bibinfo {pages} {6055} (\bibinfo {year}
  {2010})},\ \bibinfo {note} {pMID: 20931995},\ \Eprint
  {http://arxiv.org/abs/http://dx.doi.org/10.1021/nn1010914}
  {http://dx.doi.org/10.1021/nn1010914} \BibitemShut {NoStop}%
\bibitem [{\citenamefont {Chen}\ \emph {et~al.}(2008)\citenamefont {Chen},
  \citenamefont {Park}, \citenamefont {Boudouris}, \citenamefont {Horng},
  \citenamefont {Geng}, \citenamefont {Girit}, \citenamefont {Zettl},
  \citenamefont {Crommie}, \citenamefont {Segalman}, \citenamefont {Louie},\
  and\ \citenamefont {Wang}}]{Chen11}%
  \BibitemOpen
  \bibfield  {author} {\bibinfo {author} {\bibfnamefont {C.-F.}\ \bibnamefont
  {Chen}}, \bibinfo {author} {\bibfnamefont {C.-H.}\ \bibnamefont {Park}},
  \bibinfo {author} {\bibfnamefont {B.~W.}\ \bibnamefont {Boudouris}}, \bibinfo
  {author} {\bibfnamefont {J.}~\bibnamefont {Horng}}, \bibinfo {author}
  {\bibfnamefont {B.}~\bibnamefont {Geng}}, \bibinfo {author} {\bibfnamefont
  {C.}~\bibnamefont {Girit}}, \bibinfo {author} {\bibfnamefont
  {A.}~\bibnamefont {Zettl}}, \bibinfo {author} {\bibfnamefont {M.~F.}\
  \bibnamefont {Crommie}}, \bibinfo {author} {\bibfnamefont {R.~A.}\
  \bibnamefont {Segalman}}, \bibinfo {author} {\bibfnamefont {S.~G.}\
  \bibnamefont {Louie}}, \ and\ \bibinfo {author} {\bibfnamefont
  {F.}~\bibnamefont {Wang}},\ }\href {\doibase 10.1038/nnano.2008.67}
  {\bibfield  {journal} {\bibinfo  {journal} {Nature Nanotech.}\ }\textbf
  {\bibinfo {volume} {3}},\ \bibinfo {pages} {210} (\bibinfo {year}
  {2008})}\BibitemShut {NoStop}%
\bibitem [{\citenamefont {Basko}(2009)}]{Basko09}%
  \BibitemOpen
  \bibfield  {author} {\bibinfo {author} {\bibfnamefont {D.~M.}\ \bibnamefont
  {Basko}},\ }\href {http://stacks.iop.org/1367-2630/11/i=9/a=095011}
  {\bibfield  {journal} {\bibinfo  {journal} {New Journal of Physics}\ }\textbf
  {\bibinfo {volume} {11}},\ \bibinfo {pages} {095011} (\bibinfo {year}
  {2009})}\BibitemShut {NoStop}%
\bibitem [{\citenamefont {Zhao}\ \emph {et~al.}(2011)\citenamefont {Zhao},
  \citenamefont {Tan}, \citenamefont {Liu},\ and\ \citenamefont
  {Ferrari}}]{Zhao11}%
  \BibitemOpen
  \bibfield  {author} {\bibinfo {author} {\bibfnamefont {W.}~\bibnamefont
  {Zhao}}, \bibinfo {author} {\bibfnamefont {P.~H.}\ \bibnamefont {Tan}},
  \bibinfo {author} {\bibfnamefont {J.}~\bibnamefont {Liu}}, \ and\ \bibinfo
  {author} {\bibfnamefont {A.~C.}\ \bibnamefont {Ferrari}},\ }\href {\doibase
  10.1021/ja110939a} {\bibfield  {journal} {\bibinfo  {journal} {Journal of the
  American Chemical Society}\ }\textbf {\bibinfo {volume} {133}},\ \bibinfo
  {pages} {5941} (\bibinfo {year} {2011})},\ \bibinfo {note} {pMID: 21434632},\
  \Eprint {http://arxiv.org/abs/http://dx.doi.org/10.1021/ja110939a}
  {http://dx.doi.org/10.1021/ja110939a} \BibitemShut {NoStop}%
\end{thebibliography}%


%

\end{document}